\documentclass[preprint,12pt]{elsarticle}
\usepackage{amsmath,amssymb}
\usepackage{graphicx}
\usepackage{booktabs}
\usepackage{array}
\usepackage{microtype}

\usepackage[ruled,vlined,linesnumbered]{algorithm2e}
\SetAlgoNlRelativeSize{-1}  
\SetKwInput{KwData}{Data}
\SetKwInput{KwResult}{Result}

\usepackage{lineno}
\usepackage{hyperref}
\usepackage[capitalise]{cleveref}
\crefname{algocf}{Algorithm}{Algorithms}
\Crefname{algocf}{Algorithm}{Algorithms}

\newcommand{\nims}{Research Center for Macromolecules and Biomaterials, National Institute for Materials Science (NIMS), 1-1 Namiki, Tsukuba, Ibaraki 305-0044, Japan}
\newcommand{\tsukuba}{Materials Science and Engineering, Graduate School of Pure and Applied Science, University of Tsukuba, 1-1-1 Tennodai, Tsukuba, Ibaraki 305-8571, Japan}

\makeatletter
\def\ps@pprintTitle{%
  \let\@oddhead\@empty
  \let\@evenhead\@empty
  \def\@oddfoot{\footnotesize\parbox{\textwidth}{\centering
    Accepted manuscript. Published in \emph{Materials Today Communications} (2026) 116798,
    \url{https://doi.org/10.1016/j.mtcomm.2026.116798}.\\
    \textcopyright\ 2026 The Authors. Distributed under the CC BY 4.0 license.}}%
  \let\@evenfoot\@oddfoot}
\makeatother

\begin{document}

\begin{frontmatter}

\title{Through-thickness diffusion and viscoelastic relaxation
       in coated-beam chemical sensors}

\author[nims,tsukuba]{YingCheng Zhou\corref{cor1}}
\ead{ZHOU.Yingcheng@nims.go.jp}
\author[nims]{Kosuke Minami}
\author[nims,tsukuba]{Genki Yoshikawa}
\ead{YOSHIKAWA.Genki@nims.go.jp}
\cortext[cor1]{Corresponding author}
\address[nims]{\nims}
\address[tsukuba]{\tsukuba}

\begin{abstract}
Transient signals from sorption-based chemical sensors reflect both analyte uptake and the transduction mechanism.
In static-mode nanomechanical sensors, sorption-induced stress and deformation generate the signal.
Models that assume uniform uptake cannot resolve the concentration and stress gradients that develop through the coating thickness.
Here, we examine how diffusion and mechanical relaxation shape this response in a bilayer beam with a linear viscoelastic sorptive coating.
We resolve the through-thickness fields with a reduced finite-element formulation that needs one linear mechanical solve per time step.
As the ratio of diffusion to relaxation time increases, the curvature changes from a transient overshoot of its steady state to a monotonic rise.
The predictions agree with a published analytical solution in the thin-coating limit and with an independent two-dimensional continuum model for thicker coatings.
For the baseline coating parameters, the thickness that maximizes the steady-state signal exceeds the largest thickness with an overshoot above 2\%.
A map over the controlling dimensionless groups shows that a larger ratio of unrelaxed to relaxed modulus sustains the overshoot to thicker coatings and can reverse this order.
Finite surface mass-transfer resistance leaves the overshoot largely unchanged as long as uptake remains faster than mechanical relaxation.
In contrast, a gradual rise in surface concentration suppresses it once its rise time approaches the relaxation time.
The predicted regimes are consistent with published cantilever measurements.
These results guide coating selection and control of concentration rise time in coated cantilever sensors, and motivate geometry-specific analyses of other stress-based nanomechanical sensors.
\end{abstract}

\begin{keyword}
Diffusion-induced bending \sep Viscoelasticity \sep Microcantilever sensor \sep Finite element method \sep Standard linear solid
\end{keyword}

\end{frontmatter}


\section{Introduction}\label{sec:intro}

Chemical sensors support applications ranging from biomolecular detection to artificial olfaction by converting analyte-material interactions into measurable signals~\cite{fritz2000a,yoshikawa2011c,minami2022b}. Nanomechanical sensors encompass a variety of device architectures, including cantilever sensors~\cite{fritz2000a} and the Membrane-type Surface stress Sensor (MSS)~\cite{yoshikawa2011c,minami2022b}. Their static-mode operation converts analyte-induced stress and deformation into a measurable signal~\cite{fritz2000a,yoshikawa2011c,minami2022b}. Receptor layers for these sensors can be formed from diverse materials, including polymers and functionalized nanoparticles~\cite{minami2022b}. The present study focuses on coated cantilever sensors operated in static mode. A coated cantilever consists of a sorptive receptor layer on a stiff substrate such as silicon. Analyte uptake can swell this layer and bend the beam. For viscoelastic coatings, diffusion and mechanical relaxation jointly shape the time-dependent bending response. This transient carries information on sorption kinetics and analyte identity~\cite{minami2021b,minami2024b}. For example, sorption kinetic parameters extracted from the transient response of viscoelastic receptor layers have been used to discriminate twenty odorant vapors, including an aging-related aldehyde, and to quantify their concentrations~\cite{minami2024b}. Such transients are exploited in the detection of volatile organic compounds and in artificial olfaction with sensor arrays~\cite{minami2022b}. Predicting the transient is therefore important for interpreting signals acquired before steady state and for sensitivity optimization and quantitative calibration.

Existing models capture only part of this behavior. Some assume that the coating takes up analyte uniformly, following pseudo-first-order kinetics~\cite{minami2021b,minami2024b}. Others include viscoelastic relaxation of the coating and thereby reproduce a curvature \emph{overshoot}, a transient rise above the steady state~\cite{wenzel2008c,heinrich2009b}. Neither resolves the concentration gradient that Fickian diffusion sets up across the coating. Neither therefore captures how the moving diffusion front interacts with the relaxation of stress generated at different depths, and it is this interaction that shapes the full transient. With the same constitutive law and the parameters examined here, a first-order uniform-uptake model gives nearly the same overshoot ratio, but its time course differs from the spatially resolved response by up to $0.40$ of the steady-state signal. When the total uptake history is matched, neglecting the through-thickness gradient alone still changes the time course by up to $0.16$ of the steady-state signal (\cref{sec:discussion}).

Yang~\cite{yang2017a} derived an analytical solution for diffusion-induced bending of viscoelastic beams that does retain the concentration profile, giving the transient curvature in integral form. For a thin soft coating on a stiff substrate, it reduces to a compact zeroth-order expression (Eq.~77 of Ref.~\cite{yang2017a}), which we adopt as our benchmark. However, this particular expression is valid only in the thin-film limit and assumes that the surface concentration jumps instantly to its final value. It therefore cannot describe coatings of intermediate thickness, a finite rate of mass transfer across the surface, or gradual exposure to the analyte.

Resolving the through-thickness fields numerically requires accounting for transverse beam response, through-thickness diffusion, and a hereditary viscoelastic constitutive law. We develop a reduced finite-element formulation in which the curvature-dependent part of the viscoelastic overstress is absorbed into an effective beam stiffness. After each diffusion update, the mechanical response then requires only one linear solve per time step, without inner coupling iterations. Compared with a full two- or three-dimensional model in a general-purpose multiphysics package, this formulation keeps the cost per step low enough to make the parameter sweeps of \cref{sec:crossover,sec:map,sec:bc} practical. On the same computer, a single transient of the cross-check cases takes $0.2$ to $1.2$\,s, compared with about $3.5$ minutes for the two-dimensional continuum model used for cross-verification. The solver is verified component by component against standard solutions. It is cross-verified against Yang's analytical transient across the tested transition from overshoot to monotonic rise, and against an independent two-dimensional continuum model in a commercial finite-element package.

The solver is used to examine two design questions beyond the zeroth-order transient benchmark. First, the coating thickness that gives the largest steady-state signal and the largest thickness at which an overshoot of more than $2\%$ appears are governed by different factors. The former depends on bilayer geometry and relative stiffness, whereas the latter also depends on diffusion and relaxation timescales. These thicknesses do not coincide for the baseline parameters. Second, finite surface mass-transfer resistance preserves the overshoot as long as coating uptake remains faster than relaxation. In contrast, a gradual rise in surface concentration can suppress it when its rise time approaches the relaxation time. A map over the controlling dimensionless groups then shows when the overshoot appears and when the static optimum and a pronounced overshoot can coexist.

More broadly, the bilayer analysis illustrates how nonuniform uptake and finite coating stiffness can affect the interpretation of transient bending signals beyond uniform-uptake and thin-coating descriptions. It also motivates geometry-specific analyses of other stress-based nanomechanical sensors such as MSS, using appropriate constitutive and signal-transduction relations. For coated chemical sensors using other transduction mechanisms, the broader implication is not to apply the present bending model directly but to examine how spatially nonuniform uptake influences their own signal-generating processes. Such extensions require independent analytical or experimental validation and are beyond the present scope.

\section{Mathematical model}\label{sec:model}

The bilayer geometry is shown in \cref{fig:mesh}. The viscoelastic coating (layer~1, thickness $h_1$) is characterized by its relaxed and unrelaxed moduli $E_R$ and $E_U$ and its relaxation time $\tau_r$. The elastic substrate (layer~2, thickness $h_2$) is characterized by its modulus $E_2$. Our transient benchmark is Yang's analytical solution~\cite{yang2017a}, which uses a different convention for the viscoelastic law. \Cref{tab:symbols} maps between the two notations.

\begin{table}[ht]
\centering
\caption{Mapping of the viscoelastic and chemical symbols used here onto the notation of the analytical benchmark of Yang~\cite{yang2017a}, whose Maxwell-type standard linear solid uses spring moduli $Y_1$, $Y_2$ and dashpot viscosity $\eta$, with chemical strain entering as the linear eigenstrain $\tfrac13\Omega c$. The coating and substrate thicknesses $h_1$, $h_2$ and the substrate modulus $E_2$ are common to both notations.}
\label{tab:symbols}
\begin{tabular}{lll}
\toprule
Quantity & This paper & Yang~\cite{yang2017a} \\
\midrule
Unrelaxed (instantaneous) modulus & $E_U$ & $Y_1 + Y_2$ \\
Relaxed (equilibrium) modulus  & $E_R$ & $Y_1$ \\
Maxwell-arm stiffness      & $E_U - E_R$ & $Y_2$ \\
Viscoelastic relaxation time & $\tau_r$ & $\eta/Y_2$ \\
Chemical eigenstrain coefficient & $\lambda$ & $\Omega/3$ \\
\bottomrule
\end{tabular}
\end{table}

\subsection{Bilayer cross-section}

The neutral axis position, measured from the substrate bottom, is
\begin{equation}
  \bar{z}_n = \frac{E_2 h_2 (h_2/2) + E_R h_1 (h_2 + h_1/2)}{E_2 h_2 + E_R h_1}.
\end{equation}
All subsequent $z$-coordinates are measured from this neutral axis.
The integration domains for the two layers are
\begin{equation}
  A_2\colon z \in [-\bar{z}_n,\; h_2 - \bar{z}_n], \qquad
  A_1\colon z \in [h_2 - \bar{z}_n,\; h_2 + h_1 - \bar{z}_n].
\end{equation}
The section integrals are
\begin{equation}
  I_2 = b \int_{A_2} z^2 \,dz, \quad
  I_1 = b \int_{A_1} z^2 \,dz, \quad
  S_1 = b \int_{A_1} z \,dz, \quad
  EI_\mathrm{eff} = E_2 I_2 + E_R I_1,
\end{equation}
where $b$ is the beam width.

\subsection{Governing equations}

\paragraph{Kinematics}
The cantilever is modeled as an Euler--Bernoulli beam of length $L$.
The plane-sections-remain-plane assumption gives the axial strain at distance $z$ from the neutral axis:
\begin{equation}
  \varepsilon(z,t) = -z\,\kappa(x,t),
  \qquad
  \kappa = \frac{\partial^2 w}{\partial x^2},
  \label{eq:kinematics}
\end{equation}
where $w(x,t)$ is the transverse deflection and $\kappa$ the beam curvature. The reference axis is fixed at the $E_R$-weighted neutral axis, and the uniform axial strain of the bilayer is not carried as an unknown. This is exact in the steady state, where bending about the modulus-weighted axis decouples from extension, and it is accurate when the substrate dominates the axial stiffness even with the unrelaxed coating modulus ($E_U h_1 \ll E_2 h_2$). For the thickness study of \cref{sec:crossover}, $E_U h_1/(E_2 h_2) \le 0.1$. The modulus-ratio study of \cref{sec:map} extends this ratio to $0.5$ at $E_U/E_R = 10$ and $m = 10$. \ref{app:axial} compares this formulation with a bilayer calculation that carries the uniform axial strain as an additional unknown and enforces both axial force and moment balance. The two agree exactly in the steady state. For the thickness study, the overshoot ratios differ by less than $0.3\%$, and the fixed axis overestimates the instantaneous unrelaxed response by $0.5\%$, $1.4\%$, and $3.0\%$ at $m = 1$, $4.2$, and $10$. Over the range of \cref{sec:map}, the differences grow to $2.7\%$ in the overshoot ratio and $13.5\%$ in the instantaneous response at $E_U/E_R = 10$. The overshoot threshold changes by at most $1.1\%$, and the modulus ratio at which it reaches the static optimum moves from about $9.3$ to $9.5$. The two-dimensional continuum model of \cref{sec:xcheck}, which imposes no beam kinematics, is consistent with these bounds.

\paragraph{Chemical eigenstrain}
Diffusion of analyte into the viscoelastic coating induces a chemical eigenstrain proportional to the local concentration:
\begin{equation}
  \varepsilon_c(z,t) = \lambda\,C(z,t).
  \label{eq:eigenstrain}
\end{equation}
The mechanical strain driving the constitutive response is the total strain minus the eigenstrain:
\begin{equation}
  \varepsilon_\mathrm{mech} = \varepsilon - \varepsilon_c = -z\,\kappa - \lambda\,C.
  \label{eq:eps_mech}
\end{equation}

\paragraph{Constitutive laws}
The elastic substrate obeys Hooke's law:
\begin{equation}
  \sigma_2 = E_2\,\varepsilon = -E_2\,z\,\kappa.
  \label{eq:stress_sub}
\end{equation}
The viscoelastic coating is described by a standard linear solid (SLS), which consists of an equilibrium spring $E_R$ in parallel with a Maxwell element (spring $E_U - E_R$ and dashpot with time constant $\tau_r$).
The relaxation function is $G(t) = E_R + (E_U - E_R)\,e^{-t/\tau_r}$, decaying from the unrelaxed (instantaneous) modulus $E_U$ at $t = 0$ to the relaxed (long-time equilibrium) modulus $E_R$.
The stress--strain relation in hereditary integral form,
$\sigma_1(t) = \int_0^t G(t-s)\,\dot{\varepsilon}_\mathrm{mech}(s)\,ds$,
can be split into an equilibrium part and a memory part:
\begin{equation}
  \sigma_1(z,t) = E_R\,\varepsilon_\mathrm{mech}(z,t) + q(z,t),
  \label{eq:stress}
\end{equation}
where $q = \int_0^t (E_U - E_R)\,e^{-(t-s)/\tau_r}\,\dot{\varepsilon}_\mathrm{mech}(s)\,ds$ is the overstress carried by the Maxwell arm.
Differentiating this convolution shows that $q$ satisfies the first-order ODE
\begin{equation}
  \dot{q} + \frac{q}{\tau_r} = (E_U - E_R)\,\dot{\varepsilon}_\mathrm{mech}.
  \label{eq:sls}
\end{equation}
Under a step strain $\varepsilon_\mathrm{mech} = \varepsilon_0$, this gives $q(0^+) = (E_U - E_R)\varepsilon_0$ and $q(t) = (E_U - E_R)\varepsilon_0\,e^{-t/\tau_r}$, so $\sigma_1(t) = G(t)\,\varepsilon_0$, recovering the relaxation function.

\paragraph{Bending moment}
The generalized bending moment is $M = -b\!\int\!\sigma\,z\,dz$, where the integral extends over both layers.
Substituting the stress expressions for each layer:
\begin{align}
  M &= -b\!\int_{A_2}\!\sigma_2\,z\,dz - b\!\int_{A_1}\!\sigma_1\,z\,dz \nonumber\\
    &= b\!\int_{A_2}\!E_2\,z^2\,\kappa\,dz
       + b\!\int_{A_1}\!E_R\,z^2\,\kappa\,dz
       + b\!\int_{A_1}\!\lambda E_R\,C\,z\,dz
       - b\!\int_{A_1}\!q\,z\,dz \nonumber\\
    &= \underbrace{(E_2 I_2 + E_R I_1)}_{EI_\mathrm{eff}}\,\kappa
       + \underbrace{\lambda E_R S_C}_{M_C}
       - \underbrace{b\!\int_{A_1}\!q\,z\,dz}_{M_q},
  \label{eq:moment_derivation}
\end{align}
where $S_C = b\!\int_{A_1}\!C\,z\,dz$ is the concentration first moment and $M_q = b\!\int_{A_1}\!q\,z\,dz$.
Thus the total bending moment is
\begin{equation}
  M = EI_\mathrm{eff}\,\kappa + M_C - M_q.
  \label{eq:moment_total}
\end{equation}
The effective stiffness $EI_\mathrm{eff}$ uses the relaxed modulus $E_R$, and the additional stiffness from the Maxwell arm enters through $M_q$.
At $t = 0^+$, $q$ carries the full overstress, so $-M_q$ adds to the bending stiffness and the instantaneous response corresponds to the unrelaxed modulus $E_U$.
As $q$ relaxes toward zero, $M_q$ vanishes, and the moment--curvature relation approaches that of the relaxed elastic bilayer.
For spatially uniform concentration, $S_C = C \cdot S_1$, so $M_C = \lambda E_R C \cdot S_1$.
In the transient case where $C(z,t)$ is non-uniform, $S_C$ must be computed by numerical integration over the diffusion mesh at each time step.

\paragraph{Weak form}
The strong form of beam equilibrium, $\rho A\,\ddot{w} + \partial^2 M/\partial x^2 = 0$, is cast in the standard Euler--Bernoulli weak form by multiplying with an admissible test function $\delta w$ and integrating by parts twice, with the boundary terms vanishing for a cantilever. Substituting the moment \cref{eq:moment_total} and $\delta\kappa = \delta w''$ gives
\begin{equation}
  \int_0^L \rho A\,\ddot{w}\,\delta w\,dx
  + \int_0^L EI_\mathrm{eff}\,\kappa\,\delta\kappa\,dx
  = \int_0^L M_q\,\delta\kappa\,dx
    -\int_0^L \lambda E_R S_C\,\delta\kappa\,dx.
  \label{eq:weak_expanded}
\end{equation}
The left-hand side yields the standard mass and stiffness matrices after FE discretization. On the right-hand side, $M_q$ contributes the viscoelastic history force and $\lambda E_R S_C$ the chemical driving force.

\paragraph{Diffusion}
The concentration field $C(z,t)$ in the coating layer satisfies Fick's second law in the thickness direction~\cite{crank1976}:
\begin{equation}
  \frac{\partial C}{\partial t} = D\frac{\partial^2 C}{\partial z^2},
  \label{eq:diffusion}
\end{equation}
with a zero-flux Neumann condition at the coating--substrate interface ($z = 0$) and a Dirichlet condition $C = C_s$ at the free surface ($z = h_1$). In the diffusion problem, $z$ is a local coordinate measured from the interface. It is shifted by $h_2 - \bar{z}_n$ onto the neutral-axis coordinate whenever the concentration moment $S_C$ is evaluated.

\paragraph{Boundary and initial conditions}
\Cref{tab:bc} collects the boundary and initial conditions used in this work. The beam is clamped at $x = 0$ and free at $x = L$. In the quasi-static sorption calculations, the chemical loading is uniform along the beam, so the moment vanishes everywhere and the curvature is uniform. The end conditions then only remove rigid-body motion, and the tip deflection is $w(L) = \kappa L^2/2$. The diffusion problem is one-dimensional in $z$. Its boundary conditions prescribe the concentration or the flux at the two faces of the coating, and diffusion sets the profile in between. No analyte crosses the coating--substrate interface, which gives a zero-flux Neumann condition at $z = 0$. At the exposed surface, the baseline Dirichlet condition prescribes a step in surface concentration to $C_s$ at $t = 0$. Here $C_s$ is the coating concentration in equilibrium with the vapor, on the same basis as $C$, and not the vapor concentration itself. \Cref{sec:bc} examines two alternatives. A Robin condition represents finite mass transfer across the surface, with an inward flux proportional to the difference between $C_s$ and the current surface concentration. An exponentially rising surface concentration represents gradual exposure. It is a time-dependent Dirichlet condition, because the surface concentration is again prescribed.

\begin{table}[ht]
\centering
\caption{Boundary and initial conditions. The Robin condition and the exponentially rising surface concentration replace the baseline step in \cref{sec:bc}. The two-dimensional cross-check of \cref{sec:xcheck} uses a traction-free bilayer segment with rigid-body motion suppressed and the same diffusion conditions.}
\label{tab:bc}
\small
\begin{tabular}{ll>{\raggedright\arraybackslash}p{0.44\textwidth}}
\toprule
Field & Location & Condition \\
\midrule
Deflection $w$ & clamped end, $x = 0$ & $w = 0$, $\partial w/\partial x = 0$ \\
 & free end, $x = L$ & $M = 0$, $\partial M/\partial x = 0$ \\
Concentration $C$ & interface, $z = 0$ & $\partial C/\partial z = 0$ \\
 & surface, $z = h_1$ & baseline step (Dirichlet), $C = C_s$ for $t > 0$ \\
 & & Robin, $D\,\partial C/\partial z = k_s(C_s - C)$ \\
 & & exponential rise (Dirichlet), $C = C_s[1 - \exp(-t/\tau_\mathrm{bc})]$ \\
Initial state & $t = 0$ & $w = 0$, $C = 0$, $q = 0$ \\
\bottomrule
\end{tabular}
\end{table}

\subsection{Analytical benchmark}\label{sec:analytical-benchmark}

For the transient verification below we use the analytical solution for diffusion-induced bending of viscoelastic beams derived by Yang~\cite{yang2017a} as an external benchmark. In the thin-film limit, where the substrate dominates both the axial and the bending stiffness even with the unrelaxed coating modulus ($E_U h_1 \ll E_2 h_2$), the neutral axis coincides with the substrate midplane, and Yang's zeroth-order result (Eq.~77) expresses the curvature through the through-thickness moments of the diffusing concentration and a hereditary (viscoelastic) convolution over time. The concentration follows the standard Fickian series with diffusion time constants $\tau_n = 4 h_1^2 / [(2n+1)^2 \pi^2 D]$, and the hereditary convolution involves the relaxation time $\tau_r$. We refer the reader to Ref.~\cite{yang2017a} for the full derivation. The present finite-element solver retains the coating contribution to bending stiffness and therefore provides an independent check of that solution wherever the two overlap, while remaining applicable, within the model assumptions stated above, where it does not.

\section{Finite element implementation}

\subsection{Spatial discretization}

The beam is discretized with Hermite cubic elements, each node carrying two degrees of freedom: transverse deflection $w$ and rotation $\theta = dw/dx$. The element stiffness $\mathbf{K}_e = (EI/L_e^3)\,\mathbf{k}_e$ and consistent mass $\mathbf{M}_e = (\rho A L_e/420)\,\mathbf{m}_e$ take the standard $4\times 4$ Euler--Bernoulli forms~\cite{hughes2000,zienkiewicz2013}.

Through-thickness diffusion in the coating is discretized with 1D linear elements ($N_z$ elements, $N_z + 1$ nodes) with element matrices
\begin{equation}
  \mathbf{M}_z = \frac{h_e}{6}\begin{bmatrix} 2 & 1 \\ 1 & 2 \end{bmatrix}, \quad
  \mathbf{K}_z = \frac{D}{h_e}\begin{bmatrix} 1 & -1 \\ -1 & 1 \end{bmatrix}.
\end{equation}

\begin{figure}[htbp]
\centering
\includegraphics[width=\textwidth]{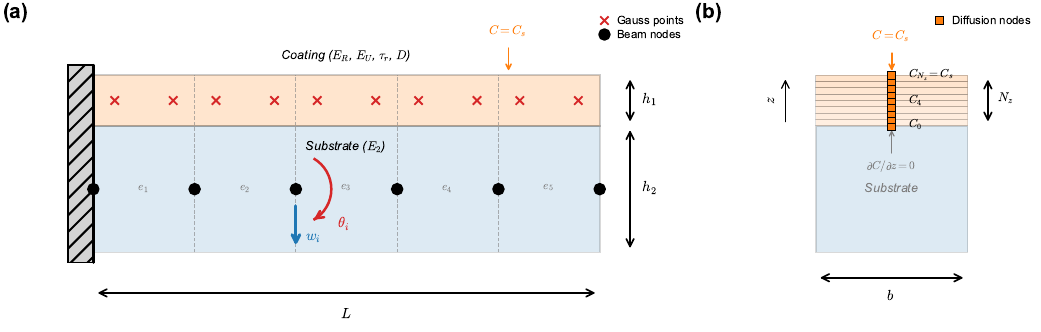}
\caption{FEM spatial discretization. (a)~Front view: Hermite beam elements with two DOFs per node ($w_i$, $\theta_i$), with crosses marking the beam Gauss integration points along the cantilever axis. (b)~Through-thickness diffusion mesh at each Gauss point: $N_z$ linear elements with $N_z + 1$ concentration nodes, a Dirichlet BC at the free surface ($z = h_1$), and a zero-flux Neumann BC at the coating--substrate interface ($z = 0$). The concentration is stored at these diffusion nodes, while the viscoelastic overstress enters the beam solve through the bending moment $M_q$, carried at the beam Gauss points of panel~(a).}
\label{fig:mesh}
\end{figure}

Cantilever boundary conditions are enforced by eliminating the two DOFs at the clamped end.
The Dirichlet diffusion boundary condition ($C = C_s$ at the top surface) is similarly enforced by DOF elimination.
Global assembly of $\mathbf{K}_e$ with $EI = EI_\mathrm{eff}$ yields the elastic beam stiffness matrix $\mathbf{K}_\mathrm{eff}$, and the global mass matrix $\mathbf{M}$ is assembled similarly from $\mathbf{M}_e$.
\Cref{fig:mesh} illustrates the spatial discretization, comprising beam elements along the cantilever axis with Gauss points in the coating layer (a) and the through-thickness diffusion mesh at each Gauss point (b).

\subsection{SLS discretization and viscoelastic stiffness}\label{sec:sls_stiffness}

\paragraph{Internal variable update}
We advance the overstress with an internal-variable recurrence in the spirit of the classical recursive treatment of viscoelastic finite elements~\cite{taylor1970,simo1987,kaliske1997,simo2000}.
\Cref{eq:sls} is a first-order linear ODE that can be integrated exactly using the integrating factor $e^{t/\tau_r}$.
Multiplying both sides:
\begin{equation}
  \frac{d}{dt}\!\bigl(q\,e^{t/\tau_r}\bigr) = (E_U - E_R)\,\dot{\varepsilon}_\mathrm{mech}\,e^{t/\tau_r}.
\end{equation}
Integrating from $t_n$ to $t_{n+1} = t_n + \Delta t$ and assuming that $\varepsilon_\mathrm{mech}(t)$ varies linearly within the step,
\begin{equation}
  \varepsilon_\mathrm{mech}(t) \approx \varepsilon^n
  + \frac{\varepsilon^{n+1} - \varepsilon^n}{\Delta t}\,(t - t_n),
  \quad t \in [t_n,\, t_{n+1}],
\end{equation}
so that $\dot{\varepsilon}_\mathrm{mech} = (\varepsilon^{n+1} - \varepsilon^n)/\Delta t$ is constant over the interval.
Evaluating the integral yields
\begin{equation}
  q^{n+1} = \underbrace{e^{-\Delta t/\tau_r}}_{\alpha}\, q^n
  + \underbrace{\frac{(E_U - E_R)\,\tau_r\,(1 - e^{-\Delta t/\tau_r})}{\Delta t}}_{\gamma_q}
    \bigl(\varepsilon_\mathrm{mech}^{n+1} - \varepsilon_\mathrm{mech}^n\bigr).
  \label{eq:sls_update}
\end{equation}
Two consistency checks confirm the correctness of \cref{eq:sls_update}:
\begin{itemize}
  \item \textbf{Constant strain} ($\varepsilon^{n+1} = \varepsilon^n$):
    the second term vanishes and $q^{n+1} = \alpha\,q^n$, i.e.\ $q$ decays exponentially toward zero, the expected relaxation behavior.
  \item \textbf{Continuous-time limit} ($\Delta t \to 0$):
    $\gamma_q \to E_U - E_R$, recovering the ODE $\dot{q} = (E_U - E_R)\,\dot{\varepsilon}_\mathrm{mech} - q/\tau_r$ in its rate form.
\end{itemize}

\paragraph{Derivation of the viscoelastic stiffness $\mathbf{K}_q$ and history moment $M_\mathrm{hist}$}
To couple the SLS update with the beam solve, we substitute
$\varepsilon_\mathrm{mech} = -z\kappa - \lambda C$ into \cref{eq:sls_update}:
\begin{equation}
  q^{n+1} = \alpha\,q^n
    + \gamma_q\bigl[-z(\kappa^{n+1}-\kappa^n) - \lambda(C^{n+1}-C^n)\bigr].
  \label{eq:q_expanded}
\end{equation}
The viscoelastic bending moment $M_q = b\!\int_{A_1}\!q\,z\,dz$ at step $n{+}1$ is obtained by multiplying \cref{eq:q_expanded} by $z$ and integrating over the coating cross-section:
\begin{equation}
  M_q^{n+1}
  = \alpha\,M_q^n
    - \gamma_q\,I_1\,(\kappa^{n+1} - \kappa^n)
    - \gamma_q\,\lambda\,\Delta S_C,
  \label{eq:Mq_update}
\end{equation}
where $I_1 = b\!\int_{A_1}\!z^2\,dz$ is the second moment of area of the coating and
$\Delta S_C = b\!\int_{A_1}\!(C^{n+1} - C^n)\,z\,dz$ captures the change in concentration moment.

The key step is to separate \cref{eq:Mq_update} into a term that depends on the unknown $\kappa^{n+1}$ and a term that depends only on known quantities at step~$n$:
\begin{equation}
  M_q^{n+1}
  = \underbrace{-\gamma_q\,I_1\,\kappa^{n+1}}_{\text{absorb into stiffness}}
  + \underbrace{\alpha\,M_q^n + \gamma_q\,I_1\,\kappa^n
    - \gamma_q\,\lambda\,\Delta S_C}_{M_\mathrm{hist}\;\text{(known)}}.
  \label{eq:Mq_split}
\end{equation}
When $M_q$ enters the weak form \cref{eq:weak_expanded} via $\int M_q\,\delta\kappa\,dx$,
the $-\gamma_q I_1\,\kappa^{n+1}$ term produces the additional stiffness matrix
\begin{equation}
  \mathbf{K}_q = \text{assemble}\bigl(\gamma_q\,I_1\bigr),
  \label{eq:Kq}
\end{equation}
which has exactly the same element structure as $\mathbf{K}_e$ (Hermite beam stiffness) but with $\gamma_q I_1$ replacing $EI$.
The total beam stiffness is
\begin{equation}
  \mathbf{K}_\mathrm{total} = \mathbf{K}_\mathrm{eff} + \mathbf{K}_q.
  \label{eq:Ktotal}
\end{equation}

The remaining history part $M_\mathrm{hist}$ is assembled into a nodal force vector via
$\mathbf{f}_\mathrm{hist} = \int \mathbf{B}_\kappa^T M_\mathrm{hist}\,dx$
and added to the right-hand side.
In the full coupled problem, the chemical moment $\lambda E_R S_C^{n+1}$ from \cref{eq:moment_total} is also moved to the right-hand side, giving the complete history force:
\begin{equation}
  M_\mathrm{hist}^\mathrm{total}
  = \alpha\,M_q^n + \gamma_q\,I_1\,\kappa^n
    - \gamma_q\,\lambda\,\Delta S_C
    - E_R\,\lambda\,S_C^{n+1}.
  \label{eq:Mhist_total}
\end{equation}

\subsection{Time integration}

The beam dynamics are integrated with the standard Newmark predictor--corrector~\cite{newmark1959} ($\beta_N = 1/4$, $\gamma_N = 1/2$; average acceleration, unconditionally stable and second-order accurate), whose effective stiffness is $\hat{\mathbf{K}} = \mathbf{M} + \beta_N\,\Delta t^2\,\mathbf{K}_\mathrm{total}$. The subscript $N$ distinguishes the Newmark parameters from $\gamma_q$ in \cref{eq:sls_update}. The sorption-response calculations (\crefrange{sec:verification}{sec:bc}) were performed in the quasi-static mechanical limit, with the mass matrix set to zero so that each Newmark step reduces to a static solve. The dynamic formulation was used separately for the vibration verification in \ref{app:components}.

Diffusion is advanced with backward Euler:
\begin{equation}
  (\mathbf{M}_z + \Delta t\,\mathbf{K}_z)\,\mathbf{C}^{n+1} = \mathbf{M}_z\,\mathbf{C}^n + \Delta t\,\mathbf{b}_z^{n+1}.
\end{equation}

\subsection{Robin boundary condition implementation}\label{sec:robin_impl}

The baseline Dirichlet condition $C(z = h_1) = C_s$ is enforced by DOF elimination.
For cases with finite surface mass transfer resistance, we replace it with a Robin condition
\begin{equation}
  D\frac{\partial C}{\partial z}\bigg|_{z=h_1} = k_s(C_s - C),
\end{equation}
characterized by the (mass-transfer) Biot number $\mathrm{Bi} = k_s h_1 / D$~\cite{crank1976}. The baseline Dirichlet condition is the $\mathrm{Bi} \to \infty$ limit of this Robin condition, recovered when surface transfer is fast compared with through-thickness diffusion.
In the finite-element discretization, this natural boundary condition adds a surface conductance term $k_s$ to the $(N_z{+}1,\, N_z{+}1)$ entry of the diffusion stiffness matrix $\mathbf{K}_z$ and a source term $k_s C_s$ to the corresponding entry of the load vector $\mathbf{b}_z$, replacing the Dirichlet DOF elimination.

For the exponentially rising surface concentration, a time-dependent Dirichlet condition, the surface value is prescribed as $C_\mathrm{top}(t) = C_s[1 - \exp(-t/\tau_\mathrm{bc})]$, with $\tau_\mathrm{bc}$ controlling the rise time.
This is implemented by updating the eliminated DOF value at each time step.

\subsection{Coupled algorithm}

The three sub-problems (diffusion, beam dynamics, viscoelastic update) are
coupled through a staggered scheme summarized in \cref{alg:coupled}.
Because the SLS term proportional to $\kappa^{n+1}$ has already been absorbed
into $\mathbf{K}_\mathrm{total}$, the mechanical solve at line~7 is a single
linear system per step, with no inner iterations required.

\begin{algorithm}[ht]
\caption{Staggered diffusion--viscoelastic--dynamics coupling}\label{alg:coupled}
\KwData{$\mathbf{d}^n,\;\mathbf{v}^n,\;\mathbf{a}^n,\;\mathbf{C}^n,\;
        M_q^n,\;\kappa^n,\;S_C^n$}
\KwResult{$\mathbf{d}^{n+1},\;\mathbf{v}^{n+1},\;\mathbf{a}^{n+1},\;
          \mathbf{C}^{n+1},\;M_q^{n+1},\;\kappa^{n+1},\;S_C^{n+1}$}
\BlankLine
\tcc{1. Diffusion sub-step (backward Euler)}
Solve $(\mathbf{M}_z + \Delta t\,\mathbf{K}_z)\,\mathbf{C}^{n+1}
       = \mathbf{M}_z\,\mathbf{C}^n + \Delta t\,\mathbf{b}_z^{n+1}$\;
\BlankLine
\tcc{2. Chemical moment update}
$S_C^{n+1} \leftarrow b\displaystyle\int C^{n+1}(z)\,z\,dz$\;
$\Delta S_C \leftarrow S_C^{n+1} - S_C^{n}$\;
\BlankLine
\tcc{3. History force assembly}
$M_\mathrm{hist} \leftarrow \alpha\, M_q^n
  + \gamma_q I_1\,\kappa^n
  - \gamma_q \lambda\,\Delta S_C
  - E_R \lambda\, S_C^{n+1}$\;
$\mathbf{f}_\mathrm{hist} \leftarrow \textsc{Assemble}(M_\mathrm{hist})$\;
\BlankLine
\tcc{4. Newmark predictor-corrector}
$\tilde{\mathbf{d}} \leftarrow \mathbf{d}^n + \Delta t\,\mathbf{v}^n
  + \tfrac{\Delta t^2}{2}(1-2\beta_N)\,\mathbf{a}^n$\;
$\mathbf{a}^{n+1} \leftarrow \hat{\mathbf{K}}^{-1}\bigl(
  \mathbf{f}_\mathrm{ext} + \mathbf{f}_\mathrm{hist}
  - \mathbf{K}_\mathrm{total}\,\tilde{\mathbf{d}}\bigr)$\;
$\mathbf{d}^{n+1} \leftarrow \tilde{\mathbf{d}}
  + \beta_N\,\Delta t^2\,\mathbf{a}^{n+1}$\;
$\mathbf{v}^{n+1} \leftarrow \mathbf{v}^n
  + \Delta t\bigl[(1-\gamma_N)\,\mathbf{a}^n
  + \gamma_N\,\mathbf{a}^{n+1}\bigr]$\;
\BlankLine
\tcc{5. Post-processing \& state update}
$\kappa^{n+1} \leftarrow \mathbf{B}_\kappa\,\mathbf{d}^{n+1}$
  \tcp*{at each Gauss point}
$M_q^{n+1} \leftarrow \alpha\, M_q^n
  - \gamma_q I_1(\kappa^{n+1}-\kappa^n)
  - \gamma_q \lambda\,\Delta S_C$\;
$S_C^{n} \leftarrow S_C^{n+1}$\;
\end{algorithm}

\section{Verification}\label{sec:verification}

\Cref{tab:common} lists the discretization and solver parameters shared across all verification cases.
Case-specific material and loading parameters are given in each subsection. The beam length and width are chosen for convenience. Under an eigenstrain that is uniform along the beam, the quasi-static curvature is independent of $L$ and $b$, so the normalized curvatures reported below carry over to microcantilever dimensions.

\begin{table}[ht]
\centering
\caption{Common simulation parameters for all verification cases.}
\label{tab:common}
\small
\begin{tabular}{ll}
\toprule
Parameter & Value \\
\midrule
Beam length $L$          & 0.2\,m \\
Beam width $b$           & 10\,mm \\
Beam elements            & 10 (Hermite cubic, 2~DOF/node) \\
Gauss points per element & 2 \\
Newmark parameters       & $\beta_N = 1/4$, $\gamma_N = 1/2$ (average acceleration) \\
Diffusion elements $N_z$ & 8 (linear, backward Euler) \\
BC enforcement           & DOF elimination (clamped end, Dirichlet diffusion) \\
\bottomrule
\end{tabular}
\end{table}

\subsection{Component verification}\label{sec:component-verify}

Before turning to the coupled problem, each physical component was verified independently against a standard reference solution. We summarize the outcomes here and give the full comparisons in \ref{app:components}.
The elastic beam reproduces the first three cantilever natural frequencies to within $0.03\%$, and first-mode free vibration conserves energy to $\sim 10^{-12}$ over ten periods, confirming the non-dissipative average-acceleration Newmark scheme (\cref{fig:freq}).
The standard-linear-solid update reproduces the analytical creep response for modulus ratios $(E_U - E_R)/E_R$ from $0.5$ to $10$, with a worst-case error that follows the expected first-order slope in the time step and stays below $2\%$ up to $\Delta t/\tau_r = 0.2$ (\cref{fig:creep}).
Through-thickness diffusion matches the Fourier-series solution to better than $0.02\,C_s$ at all times, and the resulting mass uptake $M(t)=\int_0^{h_1}C\,dz$, the quantity that drives the chemical eigenstrain, agrees with the analytical uptake to within $0.5\%$ and shows the expected Fickian $\sqrt{t}$ behavior (\cref{fig:diffusion}).
The substantive tests concern the coupled solver.

\subsection{Static bilayer coupling}

In its steady-state limit, reached under a spatially uniform eigenstrain $\varepsilon_c = \lambda C_s$, the coupled solver must reproduce the static Timoshenko bilayer curvature~\cite{timoshenko1925}
\begin{equation}
  \kappa_\mathrm{ss} = \frac{6nm(1+m)}{1 + 4nm + 6nm^2 + 4nm^3 + n^2m^4} \cdot \frac{\varepsilon_c}{h_2},
  \label{eq:timoshenko}
\end{equation}
where $m = h_1/h_2$ and $n = E_R/E_2$, and the Stoney thin-film limit $\kappa_\mathrm{Stoney} = 6n m\,\varepsilon_c / h_2$~\cite{stoney1909a}.
Parameters: $E_2 = 200$\,GPa, $E_R = 1$\,GPa ($n = 0.005$), $\lambda = 0.01$, $C_s = 1.0$\,mol/m$^3$ ($\varepsilon_c = 0.01$), $h_2 = 1$\,mm, 20~beam elements, no viscoelastic arm ($E_U = E_R$).
FEM static solutions are computed at 15 thickness ratios spanning $m = 0.005$--$50$.

\begin{figure}[htbp]
\centering
\includegraphics[width=0.55\textwidth]{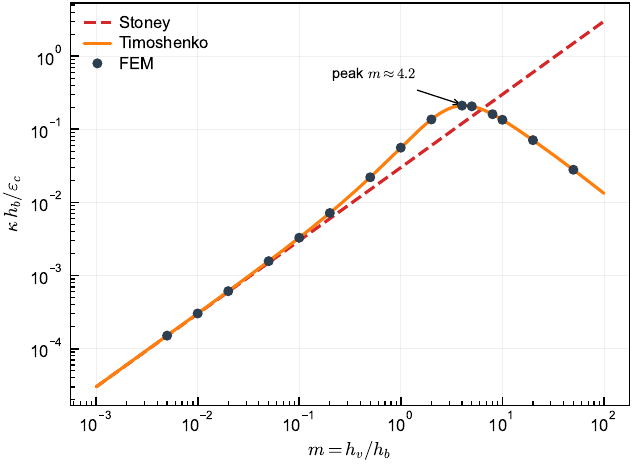}
\caption{Static bilayer coupling: normalized curvature vs.\ thickness ratio $m = h_1/h_2$ on log--log axes for Stoney (dashed), Timoshenko (solid), and FEM (circles). The FEM matches Timoshenko to within 0.5\% over four decades of $m$. The Timoshenko curvature peaks at the static optimum $m_\mathrm{opt} \approx 4.2$ and then decreases, whereas the Stoney thin-film formula diverges.}
\label{fig:static}
\end{figure}

\Cref{fig:static} compares the normalized curvature $\kappa h_2/\varepsilon_c$ over five decades of $m$.
The FEM points (circles) agree with the Timoshenko formula (solid line) to within 0.5\% across all four decades they span ($m = 0.005$--$50$), confirming the correctness of the neutral axis calculation and chemical moment assembly.
The Timoshenko curvature is \emph{non-monotonic}: it increases linearly at small $m$, reaches a maximum at $m \approx 4.2$ (where $\kappa h_2/\varepsilon_c \approx 0.21$), then decreases, approximately as $1/m$ over the range shown.
Over the thickness range shown beyond the peak, the increase in bending stiffness outweighs the increase in chemical moment, and the curvature decreases approximately as $1/h_1$. In the strict limit $m \to \infty$ at fixed $n$, \cref{eq:timoshenko} instead gives $\kappa_\mathrm{ss} h_2/\varepsilon_c \sim 6/(n m^2)$.
The Stoney formula (dashed line), by contrast, predicts an unbounded linear increase because it neglects the coating's contribution to the bending stiffness entirely. Its error relative to Timoshenko exceeds 5\% at $m > 0.054$ and 10\% at $m > 0.11$, and it cannot predict the optimum at all.
We call the location of this peak the \emph{static optimum}, $m_\mathrm{opt} \approx 4.2$. For a fixed modulus ratio, this optimum is independent of diffusion and relaxation timescales. A thicker coating raises the static signal only up to $m_\mathrm{opt}$ and lowers it beyond, so the full Timoshenko formula is essential for thick coatings. \Cref{sec:crossover} compares $m_\mathrm{opt}$ with the thickness beyond which the transient overshoot falls below $2\%$.

\subsection{Transient coupled response from overshoot to monotonic rise}\label{sec:transient}

The fully coupled solver is cross-verified against Yang's analytical transient (Eq.~77) over the complete time history. A single dimensionless group controls the shape of that history, the ratio $\tau_0/\tau_r$ of the diffusion time $\tau_0 = 4h_1^2/(\pi^2 D)$ to the viscoelastic relaxation time $\tau_r$. We hold the coating fixed and sweep the analyte diffusivity, so that one material traces the entire transition as $\tau_0/\tau_r$ varies.

The coating is thin-film ($h_1 = 1$\,$\mu$m, $h_2 = 500$\,$\mu$m, $E_2 = 170$\,GPa, $E_U = 1.7$\,GPa, $E_R = 0.85$\,GPa so $E_U/E_R = 2$, $\tau_r = 38$\,s), with $nm = 10^{-5}$ so that Yang's zeroth-order solution applies throughout. The diffusivity is tuned to give six values of $\tau_0/\tau_r$ from $0.01$ to $10$. For each, the time step $\Delta t = \min(\tau_r/20,\, \tau_0/20,\, h_1^2/(6D))$ resolves both the diffusion fill and the relaxation, and integration runs to $t_\mathrm{end} = 10\max(\tau_0,\tau_r)$.

\begin{figure}[htbp]
\centering
\includegraphics[width=0.62\textwidth]{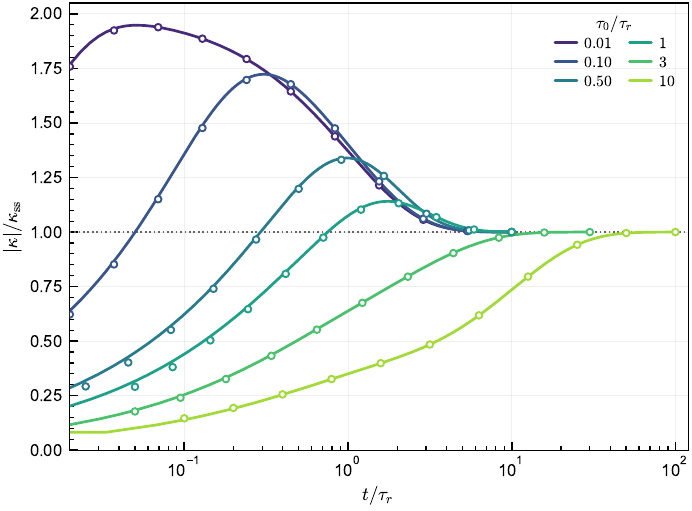}
\caption{Coupled transient verified against Yang's analytical solution (Eq.~77 of Ref.~\cite{yang2017a}, lines) versus FEM (open circles), for one thin-film coating and six analyte diffusivities spanning $\tau_0/\tau_r$ from $0.01$ to $10$, with time in units of the relaxation time $\tau_r$. The response morphs continuously from a pronounced overshoot ($1.95\,\kappa_\mathrm{ss}$ at $\tau_0/\tau_r = 0.01$) to a monotonic rise ($\tau_0/\tau_r \gtrsim 3$). The FEM reproduces Yang's transient at every point, to within $0.03\,\kappa_\mathrm{ss}$.}
\label{fig:transient}
\end{figure}

\Cref{fig:transient} shows the normalized curvature $|\kappa|/\kappa_\mathrm{ss}$ for the whole family. The FEM (markers) tracks Yang's solution (lines) at every value of $\tau_0/\tau_r$, with a worst-case discrepancy below $0.03\,\kappa_\mathrm{ss}$.
The transient shape morphs continuously with the ratio. When diffusion is fast ($\tau_0/\tau_r \ll 1$), the coating absorbs analyte nearly uniformly before significant relaxation occurs, so the initial curvature reflects the unrelaxed modulus $E_U$. As the material then relaxes toward $E_R$, the curvature \emph{overshoots} its steady state, reaching $1.95\,\kappa_\mathrm{ss}$ at $\tau_0/\tau_r = 0.01$. This peak approaches the modulus ratio $E_U/E_R = 2$ as diffusion becomes instantaneous, so a more strongly relaxing coating gives a larger overshoot.
As $\tau_0/\tau_r$ increases, the overshoot shrinks. Once diffusion is slow enough ($\tau_0/\tau_r \gtrsim 3$), each depth relaxes as fast as it fills, so the overstress never accumulates and the curvature rises \emph{monotonically} to the same steady state.

The overshoot is not a numerical curiosity. Its peak, here nearly twice the steady-state deflection, arrives long before the beam settles, so it makes the sensor respond faster and more strongly than the steady-state value alone would suggest. It also introduces an interpretive risk, since a peak mistaken for the steady-state response overstates the absorbed amount. Whether the overshoot is present therefore matters for both sensitivity and calibration, a question we take up in \cref{sec:crossover}.

\subsection{Discretization convergence and independent continuum cross-check}\label{sec:xcheck}

The coupled solver has three discretization parameters, the number of through-thickness diffusion elements $N_z$, the number of beam elements, and the time step. \Cref{tab:convergence} refines each in turn for the thickness study of \cref{sec:crossover}. It reports the overshoot ratio at $m = 0.5$ and $m = 2$ and the overshoot threshold $m_\mathrm{os}$, the outputs most sensitive to discretization. The steady-state signal is unchanged to five significant digits in all cases. The number of beam elements has no effect, because the chemical loading and hence the curvature are uniform along the beam, which the cubic Hermite elements represent exactly. The time step is the dominant source of error and converges at first order, as expected from the backward Euler diffusion step. With the settings used in this work, the overshoot ratio lies within $0.003$ and $m_\mathrm{os}$ within $0.004$ of the finest discretization tested. Here and below, $m_\mathrm{os}$ is found by solving $\kappa_\mathrm{peak}/\kappa_\mathrm{ss} = 1.02$ for $m$ (\cref{sec:crossover}).

\begin{table}[ht]
\centering
\caption{Convergence of the coupled solver for the thickness study of \cref{sec:crossover}. One parameter is refined at a time from the settings used in this work (bold), $N_z = 8$, 10 beam elements, and $\Delta t = s\min(\tau_r/5,\,h_1^2/6D)$ with $s = 0.1$.}
\label{tab:convergence}
\small
\begin{tabular}{llccc}
\toprule
 & & \multicolumn{2}{c}{$\kappa_\mathrm{peak}/\kappa_\mathrm{ss}$} & \\
\cmidrule(lr){3-4}
Refined parameter & Value & $m = 0.5$ & $m = 2$ & $m_\mathrm{os}$ \\
\midrule
Diffusion elements $N_z$ & 4 & 1.7165 & 1.0269 & 2.046 \\
 & \textbf{8} & \textbf{1.7151} & \textbf{1.0258} & \textbf{2.039} \\
 & 16 & 1.7148 & 1.0255 & 2.037 \\
 & 32 & 1.7147 & 1.0254 & 2.036 \\
\midrule
Beam elements & 5, 10, 20, 40 & 1.7151 & 1.0258 & 2.039 \\
\midrule
Time-step factor $s$ & 0.5 & 1.7043 & 1.0226 & 2.018 \\
 & 0.2 & 1.7123 & 1.0249 & 2.033 \\
 & \textbf{0.1} & \textbf{1.7151} & \textbf{1.0258} & \textbf{2.039} \\
 & 0.05 & 1.7166 & 1.0262 & 2.041 \\
 & 0.02 & 1.7175 & 1.0264 & 2.043 \\
\bottomrule
\end{tabular}
\end{table}

As an independent check that shares neither the beam kinematics nor the code, we solved the same problem as a two-dimensional plane-stress continuum in a commercial finite-element package (COMSOL Multiphysics 6.1). The bilayer segment, of length $20(h_1 + h_2)$, is free of tractions, with rigid-body motion suppressed. Fickian diffusion is solved in the coating with the conditions of \cref{tab:bc}. The standard linear solid acts on the axial stress through one internal strain variable, which reproduces \cref{eq:stress,eq:sls} for a vanishing Poisson ratio. The continuum model imposes no neutral axis and satisfies axial force and moment balance as part of the solution. Its curvature is the least-squares slope of the axial strain through the substrate over the central half of the segment. Quadratic elements were used on a mapped mesh with 16 elements through the coating, 8 through the substrate, and 160 along the length. A twofold refinement changed the normalized response by less than $0.002\,\kappa_\mathrm{ss}$ after $t = 0.05\tau_r$.

\begin{figure}[htbp]
\centering
\includegraphics[width=\textwidth]{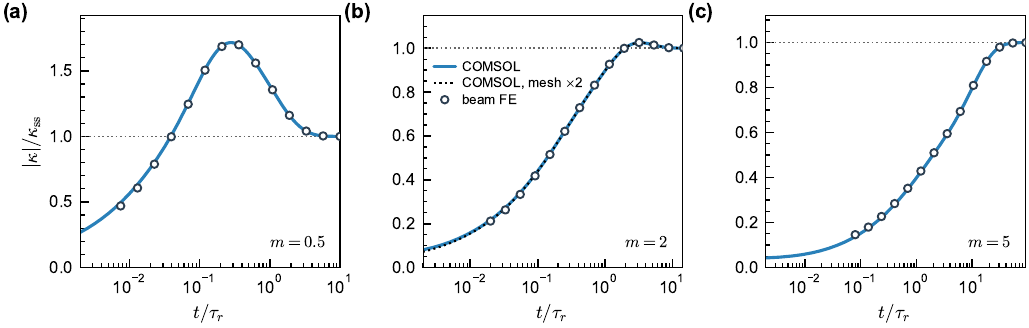}
\caption{Present beam finite-element model (open circles) compared with an independent two-dimensional plane-stress continuum model solved in COMSOL Multiphysics (lines), for three coatings of the thickness study of \cref{sec:crossover}, with curvature normalized by the steady state. (a)~$m = 0.5$, pronounced overshoot. (b)~$m = 2$, near the overshoot threshold, with the continuum result on a twofold refined mesh (dotted). (c)~$m = 5$, beyond the static optimum. The continuum model uses no beam kinematics and satisfies axial force and moment balance as part of the solution.}
\label{fig:xcheck}
\end{figure}

\Cref{fig:xcheck} compares the two models for three coatings of the thickness study, a pronounced overshoot ($m = 0.5$), a case near the overshoot threshold ($m = 2$), and a monotonic rise beyond the static optimum ($m = 5$). The steady-state curvatures agree to within $0.004\%$, the overshoot ratios to within $0.0005$, and the peak times to within about $2\%$. After $t = 0.05\tau_r$, the normalized transients differ by at most $0.009\,\kappa_\mathrm{ss}$ for $m \le 2$ and $0.02\,\kappa_\mathrm{ss}$ for $m = 5$. Larger differences appear only at the earliest times, when the diffusion front is steepest and both discretizations are least resolved. The reduced formulation thus reproduces the continuum response at a small fraction of its cost.

\section{Thickness dependence of the transient response}\label{sec:crossover}

Coating thickness controls two things at once, the size of the steady-state signal (\cref{fig:static}) and, through the diffusion time, whether an overshoot appears (\cref{sec:transient}). The thicknesses of practical interest lie where the bending stiffness of the coating becomes comparable to that of the substrate. The zeroth-order transient benchmark neglects the coating's contribution to bending stiffness and is therefore not applicable throughout this range. The present formulation retains that contribution and resolves diffusion through the coating.

\begin{figure}[htbp]
\centering
\includegraphics[width=0.62\textwidth]{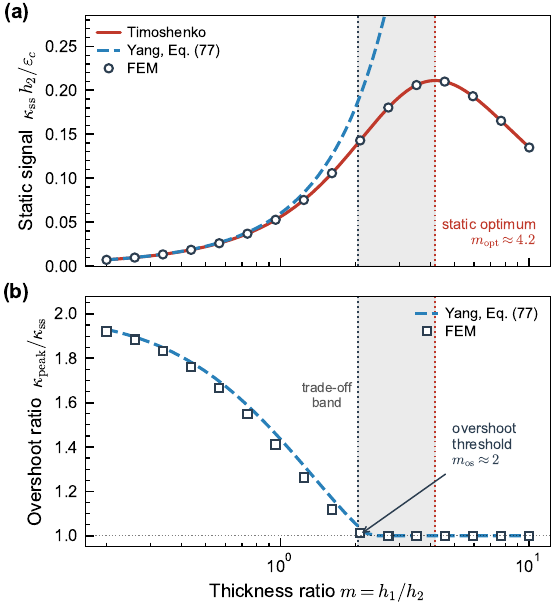}
\caption{Thickness dependence of the response from thin to thick coatings ($E_U/E_R = 2$, $\tau_r = 10$\,s). (a)~Steady-state signal $\kappa_\mathrm{ss} h_2/\varepsilon_c$ from the FEM (circles), the full Timoshenko formula (red line), and Yang's zeroth-order solution (Eq.~77, blue dashed). The FEM follows Timoshenko and peaks at the static optimum $m_\mathrm{opt} \approx 4.2$. The zeroth-order solution omits the bending stiffness of the coating and diverges once the coating thickens. (b)~Overshoot ratio $\kappa_\mathrm{peak}/\kappa_\mathrm{ss}$, which decreases toward unity and falls to $1.02$ at the overshoot threshold $m_\mathrm{os} \approx 2$. For the parameters examined, the FEM (squares) and the zeroth-order solution (dashed) give similar overshoot ratios, although their full transients differ for thicker coatings (\cref{fig:shape}). Shaded band: the trade-off region $m_\mathrm{os} < m < m_\mathrm{opt}$, where the steady-state signal is still rising but the overshoot is at most $2\%$.}
\label{fig:crossover}
\end{figure}

We fix the substrate at $h_2 = 3$\,$\mu$m and vary the coating thickness from $h_1 = 0.6$ to $30$\,$\mu$m ($m = 0.2$--$10$). The other parameters are $E_2 = 170$\,GPa, $E_R = 0.85$\,GPa, $E_U/E_R = 2$, $\tau_r = 10$\,s, and $D = 10^{-12}$\,m$^2$/s. Because the relaxation time is fixed, a thicker coating also has a longer diffusion time $\tau_0 = 4h_1^2/(\pi^2 D)$ and hence a larger ratio $\tau_0/\tau_r$. For each thickness, \cref{fig:crossover} reports two quantities, the steady-state signal $\kappa_\mathrm{ss} h_2/\varepsilon_c$ and the overshoot ratio $\kappa_\mathrm{peak}/\kappa_\mathrm{ss}$.

The steady-state signal (\cref{fig:crossover}a) follows the Timoshenko curve to within the marker size, consistent with \cref{fig:static}, and peaks at the static optimum $m_\mathrm{opt} \approx 4.2$. Yang's zeroth-order solution agrees for thin coatings. It neglects the bending stiffness of the coating, however, so it overpredicts the signal once the coating thickens and misses the peak altogether.

The overshoot ratio (\cref{fig:crossover}b) behaves differently. It is largest for thin coatings, where diffusion is fast compared with relaxation, and decreases steadily as the coating thickens and $\tau_0/\tau_r$ grows. The ratio falls to $1.02$ at $m_\mathrm{os} \approx 2$ for these parameters, which we call the \emph{overshoot threshold}. Throughout this work, $\kappa_\mathrm{peak}/\kappa_\mathrm{ss} = 1.02$ serves as an operational threshold for a resolved overshoot. Responses below it may still overshoot by less than $2\%$ and are not necessarily strictly monotonic. For the parameters examined, the zeroth-order solution and the FEM give similar overshoot ratios despite the difference in their absolute steady-state signals for thicker coatings.

\begin{figure}[htbp]
\centering
\includegraphics[width=\textwidth]{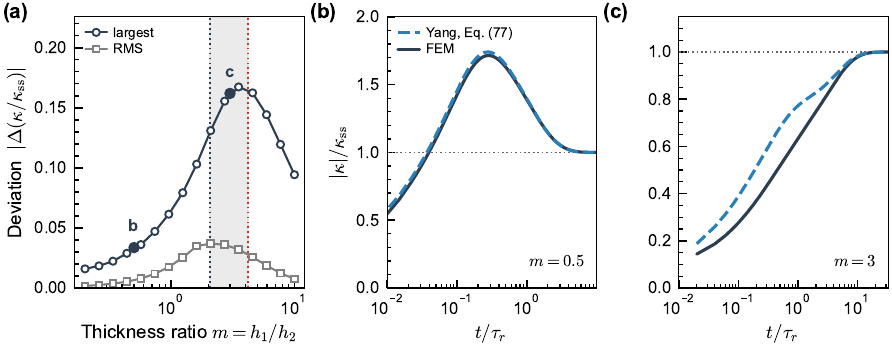}
\caption{Full transient of the FEM compared with Yang's zeroth-order solution (Eq.~77 of Ref.~\cite{yang2017a}) for the thickness sweep of \cref{fig:crossover}. Each curve is normalized by its own steady state, so the difference in static signal is removed. (a)~Largest (circles) and root-mean-square (squares) deviation of $|\kappa|/\kappa_\mathrm{ss}$ over the full response versus thickness ratio $m$. Filled circles mark the two cases shown in panels (b) and (c). Shading and dotted lines mark the trade-off band of \cref{fig:crossover}. (b)~Time course for a thin coating ($m = 0.5$), where the two solutions nearly coincide. (c)~Time course inside the trade-off band ($m = 3$), where the zeroth-order solution rises faster at early times.}
\label{fig:shape}
\end{figure}

The overshoot ratio compares only the peak with the steady state. To compare the whole time course, we normalize each transient by its own steady state, which removes the static bias of \cref{fig:crossover}a. We then compute the largest and root-mean-square deviations between the two solutions (\cref{fig:shape}a). For $m < 0.5$, the largest deviation stays below $0.03\,\kappa_\mathrm{ss}$, and the two transients nearly coincide (\cref{fig:shape}b). The deviation grows as the coating thickens and peaks inside the trade-off band, where it reaches about $0.17\,\kappa_\mathrm{ss}$ and the root-mean-square deviation stays below $0.04\,\kappa_\mathrm{ss}$. There, the zeroth-order solution rises faster than the FEM at early times (\cref{fig:shape}c). This is consistent with its omission of the coating stiffness, which at early times includes the unrelaxed modulus $E_U$ and resists bending. For still thicker coatings, diffusion becomes much slower than relaxation. The coating then stays nearly relaxed during uptake, and normalization removes most of the remaining difference. Thus, even where the zeroth-order solution predicts the overshoot ratio well, it does not reproduce the time course of the response inside the trade-off band.

The two thicknesses therefore do not coincide. For $m_\mathrm{os} < m < m_\mathrm{opt}$ (shaded in \cref{fig:crossover}), a thicker coating still increases the steady-state signal, but the overshoot has already fallen below $2\%$. No single thickness in this study gives both the largest signal and a pronounced overshoot.

The two thicknesses are set by different physics. The static optimum $m_\mathrm{opt}$ depends only on the modulus ratio $n = E_R/E_2$. The overshoot threshold $m_\mathrm{os}$ depends on the balance between diffusion and relaxation and therefore shifts with $\tau_r$, $D$, and $E_U/E_R$. Their order is thus material-specific, as the map of \cref{sec:map} shows. Locating the trade-off band requires both thicknesses. The zeroth-order model estimates $m_\mathrm{os}$ for the parameters examined, while the Timoshenko formula gives the static optimum $m_\mathrm{opt}$. The present framework evaluates the transient response while retaining finite coating stiffness, allowing these two design criteria to be compared within the same model.

\section{Conditions for the overshoot}\label{sec:map}

\begin{figure}[htbp]
\centering
\includegraphics[width=\textwidth]{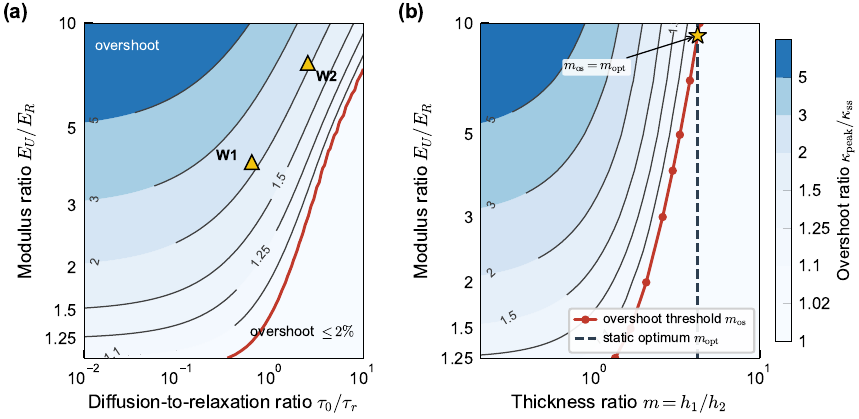}
\caption{Overshoot ratio $\kappa_\mathrm{peak}/\kappa_\mathrm{ss}$ over the controlling parameters. (a)~Thin-coating limit (Yang's solution) as a function of $\tau_0/\tau_r$ and $E_U/E_R$. The red line marks where the ratio falls to $1.02$, the operational threshold used in this work. Beyond it, any overshoot is at most $2\%$. Triangles, measured overshoots of Wenzel et al.~\cite{wenzel2008c} (W1 for dichloromethane and W2 for trichloroethylene), placed with their fitted modulus ratios and with their fitted ratio of sorption to relaxation time in place of $\tau_0/\tau_r$. (b)~Finite coatings (FEM, other parameters as in \cref{sec:crossover}) as a function of $m$ and $E_U/E_R$. The overshoot threshold $m_\mathrm{os}$ (red) moves to thicker coatings as $E_U/E_R$ increases, whereas the static optimum $m_\mathrm{opt}$ (dashed) is fixed by $E_R/E_2$. The threshold is found by solving $\kappa_\mathrm{peak}/\kappa_\mathrm{ss} = 1.02$ for $m$. The two cross near $E_U/E_R \approx 9.3$ (star).}
\label{fig:map}
\end{figure}

The thickness study varies one parameter at a time. To map where the overshoot appears and where it falls below the $2\%$ criterion, we vary the two dimensionless groups that control it, $\tau_0/\tau_r$ and $E_U/E_R$. In the thin-coating limit, the normalized response depends on these two groups only, and Yang's solution, verified in \cref{sec:transient}, gives it at low cost. \Cref{fig:map}a shows the overshoot ratio over $0.01 \le \tau_0/\tau_r \le 10$ and $1.1 \le E_U/E_R \le 10$. The overshoot is largest when diffusion is fast and the coating relaxes strongly, where the ratio approaches $E_U/E_R$. Its boundary, where the ratio falls to $1.02$, moves to slower diffusion as $E_U/E_R$ increases. The ratio falls below the $2\%$ criterion near $\tau_0/\tau_r \approx 0.7$ for $E_U/E_R = 1.25$, near $1.7$ for $E_U/E_R = 2$, near $3.8$ for $E_U/E_R = 4$, and near $9$ for $E_U/E_R = 7$. A strongly relaxing coating can therefore show an overshoot even when uptake is slower than relaxation.

For coatings of finite thickness, \cref{fig:map}b repeats the thickness study of \cref{sec:crossover} with the FEM for $E_U/E_R$ from $1.25$ to $10$. The overshoot threshold increases from $m_\mathrm{os} = 1.32$ to $4.32$ over this range. The static optimum stays at $m_\mathrm{opt} \approx 4.2$, because it depends only on $E_R/E_2$. For the diffusion and relaxation times of \cref{sec:crossover}, the two thicknesses cross near $E_U/E_R \approx 9.3$, which we obtain by solving $\kappa_\mathrm{peak}/\kappa_\mathrm{ss} = 1.02$ at $m = m_\mathrm{opt}$. This part of the map lies beyond the range for which \ref{app:axial} gives the error bounds of \cref{sec:crossover}. With full axial force balance, the crossing moves to $E_U/E_R \approx 9.5$, so its position is approximate, but the change of order within the map remains. Below this value a trade-off band exists, as in \cref{fig:crossover}. Above it, the coating that maximizes the steady-state signal still shows an overshoot. Whether the largest static signal and a pronounced transient peak can be obtained together is thus a property of the coating material, set by its modulus ratio and its diffusion and relaxation times.

\section{Boundary condition sensitivity}\label{sec:bc}

All preceding cases, like the analytical benchmark, assume that the surface concentration jumps instantly to $C_s$ (a Dirichlet condition). In practice, two effects can slow this loading. Mass transfer across the surface may be finite, and the analyte may reach the sensor gradually. We model the first with a Robin condition $D\,\partial C/\partial z = k_s(C_s - C)$ at $z = h_1$, characterized by the mass-transfer Biot number $\mathrm{Bi} = k_s h_1/D$. We model the second with an exponentially rising surface concentration, a time-dependent Dirichlet condition $C_\mathrm{top}(t) = C_s[1 - \exp(-t/\tau_\mathrm{bc})]$ with rise time $\tau_\mathrm{bc}$. \Cref{sec:robin_impl} describes the implementation.

\begin{figure}[htbp]
\centering
\includegraphics[width=\textwidth]{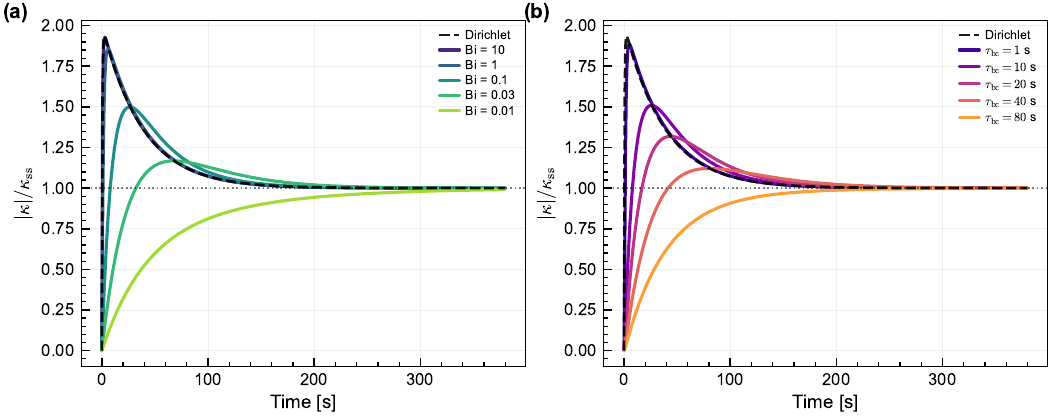}
\caption{Effect of the surface boundary condition on the overshoot regime ($h_1 = 1$\,$\mu$m, $E_U/E_R = 2$, $\tau_r = 38$\,s, $\tau_0/\tau_r \approx 0.01$), with curvature normalized by the steady state. (a)~Robin condition with $\mathrm{Bi}$ from $10$ to $0.01$. The overshoot is essentially unchanged down to $\mathrm{Bi} = 1$ and vanishes near $\mathrm{Bi} \approx 0.01 \sim \tau_0/\tau_r$. (b)~Exponentially rising surface concentration (time-dependent Dirichlet condition) with rise time $\tau_\mathrm{bc}$ from $1$ to $80$\,s. The overshoot shrinks as $\tau_\mathrm{bc}$ approaches $\tau_r$ and is gone by $\tau_\mathrm{bc} \approx 2\tau_r$. Dashed black line: instantaneous Dirichlet baseline.}
\label{fig:bc_overshoot}
\end{figure}

We apply both conditions to one case from each transient regime of \cref{sec:transient}. The overshoot case has $\tau_0/\tau_r \approx 0.01$ ($h_1 = 1$\,$\mu$m, $h_2 = 500$\,$\mu$m, $E_2 = 170$\,GPa, $E_U = 0.4$\,GPa, $E_R = 0.2$\,GPa, $\tau_r = 38$\,s, $D = 10^{-12}$\,m$^2$/s), and the monotonic case has $\tau_0/\tau_r \approx 4$ ($h_1 = 100$\,$\mu$m, $h_2 = 1$\,mm, $E_2 = 200$\,GPa, $E_U = 3$\,GPa, $E_R = 1$\,GPa, $\tau_r = 1$\,s, $D = 10^{-9}$\,m$^2$/s). In the monotonic case, neither condition changes the character of the response. A smaller Bi or a longer $\tau_\mathrm{bc}$ only slows the approach to steady state, and the curvature stays monotonic for every value tested. The overshoot case is therefore the informative one (\cref{fig:bc_overshoot}).

Surface resistance barely affects the overshoot (\cref{fig:bc_overshoot}a). The overshoot is essentially unchanged from $\mathrm{Bi} = 10$ down to $\mathrm{Bi} = 1$ and disappears only when Bi falls to about $0.01$, where the response becomes monotonic. The reason is the same timescale separation that produces the overshoot. With surface resistance, loading the coating takes a time of order $\tau_0/\mathrm{Bi}$. The overshoot survives as long as this time stays short compared with the relaxation time $\tau_r$, so it disappears near $\mathrm{Bi} \sim \tau_0/\tau_r$. Using $\tau_0 = 4h_1^2/(\pi^2 D)$, this condition becomes, up to a factor of order one, $k_s \gtrsim h_1/\tau_r$, which is independent of the diffusivity.

A gradual rise in surface concentration acts more directly (\cref{fig:bc_overshoot}b), because it sets the loading time $\tau_\mathrm{bc}$ regardless of diffusion. The overshoot shrinks as $\tau_\mathrm{bc}$ approaches $\tau_r$ and disappears once $\tau_\mathrm{bc}$ reaches about $2\tau_r$, after which the curvature rises monotonically. This dependence on the loading rate has been observed experimentally. Shemesh et al.~\cite{shemesh2013} exposed a poly(4-vinylpyridine)-coated cantilever to deuterated ethanol at carrier-gas flow rates of $100$, $60$, and $20$\,mL/min. In their setup the vapor concentration at the coating surface approaches its final value as a first-order process, the same form as our exponentially rising surface concentration, and a lower flow rate supplies the analyte more slowly. The bending overshoot fell from about twice the steady-state value at the highest flow rate to a modest overshoot at the intermediate rate, and it was absent at the lowest rate, in line with \cref{fig:bc_overshoot}b.

Measured transfer coefficients show which of these limits applies in practice. For a coating of about one micrometer with a relaxation time of a few tens of seconds, the critical value $h_1/\tau_r$ is of order $10^{-8}$\,m/s. Measured surface mass-transfer coefficients for vapor and solvent uptake by polymer films lie between $10^{-6}$ and $10^{-4}$\,m/s~\cite{kongkanand2011,nielsen2005}, roughly two to four orders of magnitude larger. Small linear penetrants show no measurable surface resistance at all~\cite{nielsen2005}. Surface control is not universal, and it can dominate the sorption of particular systems such as water in Nafion membranes~\cite{satterfield2008}. For coating-analyte systems with $k_s$ well above $h_1/\tau_r$, the present calculations support preservation of the overshoot despite finite surface resistance. This condition must be assessed for the system of interest rather than assumed to hold universally. When surface transfer is sufficiently fast, the concentration at the sensor must also rise on a timescale well below $\tau_r$ to preserve a pronounced overshoot.

\section{Discussion}\label{sec:discussion}

The thickness and boundary-condition studies address two practical questions beyond the zeroth-order transient benchmark. For the baseline coating parameters of \cref{sec:crossover}, the coating thickness that maximizes the steady-state signal exceeds the operational overshoot threshold defined by the $2\%$ criterion. The static optimum depends on bilayer geometry and relative stiffness, while the overshoot threshold $m_\mathrm{os}$ additionally depends on transport and relaxation timescales. Their relative positions are therefore parameter-dependent. Finite surface resistance suppresses the overshoot if it makes coating uptake sufficiently slow compared with relaxation. When surface transfer remains fast, a gradual rise in concentration at the sensor can instead become the controlling limitation.

These results complement earlier work in two directions. Analytical models of viscoelastically coated cantilevers reproduce the overshoot for a prescribed history of the eigenstrain and give closed-form criteria for it~\cite{wenzel2008c,heinrich2009b}. They do not, however, resolve the diffusion front that produces this eigenstrain in practice. Coupled diffusion--stress models have been developed extensively for battery electrodes~\cite{christensen2006,bower2011,miehe2016,talukdar2024}, where the geometry and the failure modes differ from those of a slender bilayer sensor. The present work examines through-thickness diffusion and viscoelastic relaxation in a reduced bilayer-sensor model with finite coating stiffness and non-instantaneous boundary loading, using a one-way transport-to-mechanics coupling.

The advantage over uniform-uptake descriptions can be quantified directly. \Cref{tab:uniform} compares the Fickian through-thickness concentration of the present model with two uniform-uptake variants that share its constitutive law and its kinematics. All three are evaluated with the bilayer model of \ref{app:axial}, so that only the concentration field differs. In the first, the coating concentration rises as a first-order process with time constant $\tau_0$, as in pseudo-first-order and earlier viscoelastic cantilever models~\cite{minami2021b,wenzel2008c}. In the second, the uniform concentration follows the exact Fickian uptake, so that only the spatial gradient is missing. All three give overshoot ratios within $1\%$ of each other. Their time courses differ considerably, however. The first-order variant differs from the Fickian case in both the total uptake history and its distribution through the thickness. It deviates by $0.29$ to $0.40\,\kappa_\mathrm{ss}$ and delays the peak by up to $25\%$. The second variant matches the total uptake history and therefore isolates the effect of the gradient. It still deviates by $0.05\,\kappa_\mathrm{ss}$ at $m = 0.2$ and by $0.15$ to $0.16\,\kappa_\mathrm{ss}$ for $m \ge 1$, because the depth at which the analyte resides changes its moment about the neutral axis. These are differences in the predicted waveform, not measured differences in analyte identification or parameter estimation. For applications that use the shape of the transient, such as the extraction of sorption kinetic parameters for analyte discrimination~\cite{minami2024b}, a through-thickness description may therefore be important for quantitative interpretation when uptake gradients appreciably affect the bending response.

\begin{table}[ht]
\centering
\caption{Fickian through-thickness concentration of the present model compared with uniform-uptake variants (parameters of \cref{sec:crossover}). All three are computed with the bilayer model of \ref{app:axial}, so that only the concentration field differs. In the Fickian case, the overshoot ratios agree with the beam finite-element solver to within $0.001$. U1, uniform concentration rising as $C_s(1 - e^{-t/\tau_0})$. U2, uniform concentration following the exact Fickian uptake $M(t)/M_\infty$. Peak quantities are omitted where the response has no overshoot. The deviation is the largest difference of $|\kappa|/\kappa_\mathrm{ss}$ from the Fickian case over the whole response.}
\label{tab:uniform}
\small
\begin{tabular}{lccccccccc}
\toprule
 & \multicolumn{3}{c}{$\kappa_\mathrm{peak}/\kappa_\mathrm{ss}$} & \multicolumn{3}{c}{$t_\mathrm{peak}/\tau_r$} & \multicolumn{2}{c}{Deviation$/\kappa_\mathrm{ss}$} \\
\cmidrule(lr){2-4}\cmidrule(lr){5-7}\cmidrule(lr){8-9}
$m$ & Fickian & U1 & U2 & Fickian & U1 & U2 & U1 & U2 \\
\midrule
0.2 & 1.921 & 1.920 & 1.921 & 0.069 & 0.073 & 0.070 & 0.33 & 0.05 \\
0.5 & 1.714 & 1.710 & 1.713 & 0.28 & 0.31 & 0.29 & 0.38 & 0.10 \\
1   & 1.386 & 1.375 & 1.380 & 0.77 & 0.90 & 0.83 & 0.40 & 0.15 \\
2   & 1.025 & 1.021 & 1.022 & 3.32 & 4.16 & 3.78 & 0.38 & 0.16 \\
4.2 & -- & -- & -- & -- & -- & -- & 0.30 & 0.15 \\
10  & -- & -- & -- & -- & -- & -- & 0.29 & 0.15 \\
\bottomrule
\end{tabular}
\end{table}

The predicted regimes agree with reported measurements on coated cantilevers. Wenzel et al.~\cite{wenzel2008c} fitted a viscoelastic model to the measured bending of microcantilevers coated with a modified $\beta$-cyclodextrin and exposed to dichloromethane and trichloroethylene, both of which overshot their steady state. Their fitted modulus ratios, $E_U/E_R \approx 4$ and $7.7$, and ratios of sorption to relaxation time, $0.63$ and $2.5$, place both measurements inside the overshoot region of \cref{fig:map}a (triangles). In the second case uptake is slower than relaxation, yet the large modulus ratio moves the boundary to slower diffusion, so an overshoot is still expected. Shemesh et al.~\cite{shemesh2013} reported two further trends with poly(4-vinylpyridine) coatings. Heavy water and deuterated ethanol produced large overshoots where the protiated species gave little or none. Shemesh et al.\ attributed this isotope effect mainly to slower relaxation, assuming comparable diffusivities for the isotopologues rather than measuring them independently. The reported trend is therefore consistent with slower relaxation relative to uptake, that is, with a lower $\tau_0/\tau_r$. Lowering the carrier flow rate removed the overshoot, as discussed in \cref{sec:bc}. These comparisons are qualitative or semi-quantitative, because the published parameters were fitted with uniform-uptake models rather than measured independently. In \cref{fig:map}a, the fitted sorption time serves as a proxy for the diffusion time and is not an independently measured value of $\tau_0$. The comparison therefore shows consistency with the predicted overshoot regimes and trends, not an independent quantitative validation or a reproduction of the measured curvature histories. A quantitative test requires independent measurements of the diffusivity and relaxation spectrum of the coating together with the transient bending.

The coating also affects dynamic-mode operation, in which the resonance frequency is the read-out. At kilohertz frequencies the coating responds with its unrelaxed modulus $E_U$, because the oscillation period is far shorter than $\tau_r$. Its mass lowers the first bending frequency, whereas its bending stiffness raises it. For the coating of \cref{sec:crossover} with a density of $1000$\,kg/m$^3$ on silicon, the frequency falls to $0.89$ of the bare value near $m = 1.2$ and recovers the bare value near $m = 2.5$ (\ref{app:dynamic}). Coating thickness therefore affects the resonance frequency and the static curvature differently. The calculated frequency trend alone does not establish an optimum thickness for dynamic sensing. The transient read-outs differ as well. In a mass-loading-only approximation with the mechanical properties held fixed, the frequency shift follows the absorbed mass, which rises monotonically as $M(t)/M_\infty$, and shows no overshoot. The static overshoot arises from the relaxation of stress generated through the thickness, a process that the absorbed mass does not undergo. This estimate does not exclude non-monotonic frequency responses caused by sorption-induced changes in stiffness or stress, for example through plasticization or surface stress.

Several assumptions limit the present framework, and their effects can be estimated. First, transport is assumed Fickian with a constant diffusivity. Whether this holds must be assessed for each coating and analyte. In glassy polymers, coupling between penetrant transport and structural relaxation can produce anomalous or Case~II transport with sharp concentration fronts~\cite{thomas1980,song2023}. Such behavior lies outside the present model, and its effect on the overshoot cannot be inferred from the uniform-uptake comparison in \cref{tab:uniform}. Second, the coating is a single standard linear solid, the simplest viscoelastic model that produces a finite overshoot. Real polymers show a broad relaxation spectrum, which spreads the overshoot over several timescales. The formulation extends directly to a Prony series, because each additional Maxwell arm adds one overstress moment of the form of $M_q$ and one term to the effective stiffness, with no change to the single linear solve per step. Third, the formulation assumes small strains and linear Euler--Bernoulli kinematics. Sorption strains in sensing coatings usually remain below a few percent and tip deflections are far smaller than the beam length, so the linear theory applies. Large-swelling coatings such as hydrogels would require a finite-strain extension with concentration-dependent moduli. Fourth, the geometry is a one-dimensional bilayer beam. The plane-stress cross-check of \cref{sec:xcheck} supports the through-thickness kinematics. That cross-check uses a vanishing Poisson ratio, however, and does not test constraint across the width. In plate-like devices, transverse constraint and Poisson effects can change the effective stiffness and the stress distribution. Their influence on the transient response requires an appropriate plate or continuum formulation. Platforms such as MSS, with two-dimensional plate bending and non-uniform coatings, require their own models. Most importantly, the coupling runs one way. Diffusion drives stress, but stress does not feed back into transport. In polymers, mechanical relaxation can change the local free volume and hence the diffusivity. This effect underlies non-Fickian Case II transport~\cite{thomas1980} and the asymmetric absorption and desorption kinetics seen in several sorption experiments~\cite{satterfield2008}. Including stress-dependent diffusion, as in fully coupled elasto-visco-plastic swelling models~\cite{song2023}, is the most immediate extension needed to capture these phenomena.

\section{Conclusion}

The transient bending of a sorptive, viscoelastically coated beam depends on both through-thickness analyte transport and the relaxation of locally generated stress. We examined these processes with a reduced finite-element formulation that retains the coating's bending stiffness and incorporates the curvature-dependent viscoelastic overstress into an effective mechanical stiffness. After each diffusion update, the mechanical response requires the solution of one linear system without inner coupling iterations. Each component was verified separately, and the coupled solver reproduces Yang's analytical transient~\cite{yang2017a} across the tested transition from overshoot to monotonic rise. For the tested finite coatings and $t \ge 0.05\tau_r$, an independent two-dimensional continuum model and the reduced model differ by at most $0.02\,\kappa_\mathrm{ss}$ in curvature, and the solver converges to within $0.003$ in the overshoot ratio.

Within the present bilayer-beam model, the static optimum and the overshoot threshold are governed by different factors and do not coincide for the baseline parameters of the thickness study. Increasing coating thickness toward the static optimum raises the diffusion-to-relaxation time ratio enough that the overshoot falls below the $2\%$ criterion. In the boundary-condition study, the overshoot is retained when the surface-transfer loading time remains short compared with the relaxation time. This gives an order-of-magnitude condition $k_s \gg h_1/\tau_r$ in the surface-transfer-limited regime. A gradual rise in surface concentration can instead reduce the overshoot as its rise time approaches the relaxation time and remove it once the rise time reaches about twice the relaxation time. Over the controlling groups $\tau_0/\tau_r$ and $E_U/E_R$, a larger modulus ratio sustains the overshoot to slower diffusion and thicker coatings. For the diffusion and relaxation times of the thickness study, the overshoot threshold reaches the static optimum near $E_U/E_R \approx 9.3$, or $9.5$ with full axial force balance. These results identify coating thickness and concentration rise time at the sensor as joint considerations when designing and interpreting transient measurements with coated cantilever sensors.

The framework is limited to a one-dimensional bilayer beam, a linear viscoelastic constitutive law, and one-way coupling in which diffusion drives stress but not the reverse. Nevertheless, the results motivate geometry-specific assessments of spatially nonuniform uptake and mechanical relaxation in other stress-based nanomechanical sensors, including MSS. For coated sensors using other transduction mechanisms, the transferable perspective is to relate the uptake distribution to the relevant signal-generating process. The present bending model does not directly describe those signals. Extensions to stress-dependent diffusion, other nanomechanical geometries, or other sensing mechanisms will require suitable formulations and independent analytical or experimental validation.

\section*{Declaration of competing interest}

The authors declare that they have no known competing financial interests or personal relationships that could have appeared to influence the work reported in this paper.

\section*{Funding}

This work was supported by the NIMS Junior Research Program, National Institute for Materials Science (NIMS), Japan, and by the Pioneering Research Initiated by the Next Generation (SPRING) program, JST, MEXT, Japan (No.\ JPMJSP2124).

\section*{CRediT authorship contribution statement}

\textbf{YingCheng Zhou}: Conceptualization, Methodology, Software, Formal analysis, Investigation, Visualization, Writing -- original draft, Funding acquisition.
\textbf{Kosuke Minami}: Writing -- review \& editing, Supervision.
\textbf{Genki Yoshikawa}: Writing -- review \& editing, Supervision.

\section*{Data availability}

The source code for the finite-element solver, the scripts used to generate all figures, and the input script of the two-dimensional COMSOL cross-check are available at \url{https://github.com/metachow/coated-beam-fem}. All numerical data presented in this study can be reproduced using the provided code.

\section*{Declaration of generative AI and AI-assisted technologies in the manuscript preparation process}

During the preparation of this work the authors used Claude (Anthropic) in order to improve the language and structure of the manuscript text, to assist in writing the plotting and post-processing scripts and the input script of the two-dimensional cross-check model, and to assist with LaTeX typesetting. ChatGPT (OpenAI) was also used to assist with revising the framing and wording of the manuscript and checking the consistency of its presentation. All suggestions were evaluated by the authors. After using these tools, the authors reviewed and edited the content as needed and take full responsibility for the content of the published article.

\appendix

\section{Component verification against standard benchmarks}\label{app:components}

This appendix collects the component-level verification summarized in \cref{sec:component-verify}. Each physical module was tested in isolation against a standard reference solution, using the shared discretization parameters of \cref{tab:common}.

\begin{figure}[htbp]
\centering
\includegraphics[width=\textwidth]{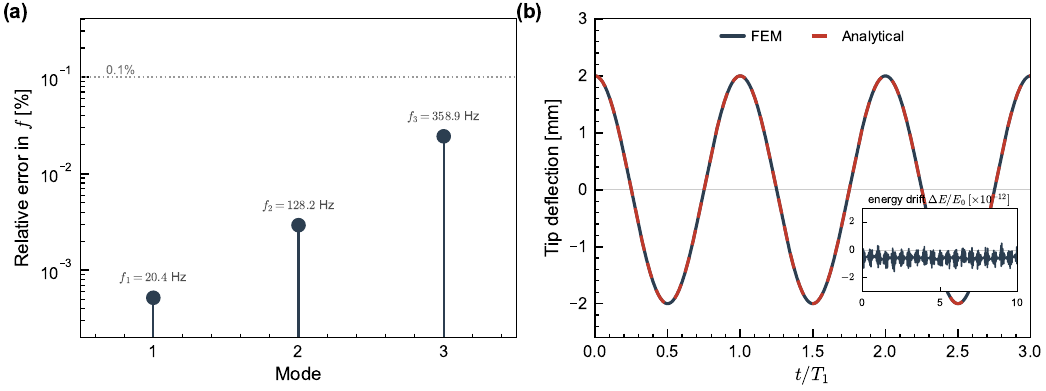}
\caption{Elastic cantilever verification. (a)~Relative error in the first three natural frequencies (FEM vs.\ analytical), all below $0.03\%$. The mode frequencies are annotated. (b)~First-mode free vibration ($w_\mathrm{tip}(0) = L/100$): FEM (solid) overlaid with the analytical cosine (dashed), shown over the first three periods. Over the full ten-period run the amplitude changes by less than $0.01\%$ and the total energy is conserved to $\sim 10^{-12}$ relative variation, confirming the non-dissipative property of the average-acceleration Newmark scheme.}
\label{fig:freq}
\end{figure}

\paragraph{Elastic beam} The first three cantilever natural frequencies of a uniform elastic beam ($h_2 = 1$\,mm, $E_2 = 200$\,GPa, $\rho_2 = 7800$\,kg/m$^3$; obtained from the bilayer formulation by setting $h_1 = 0$) are compared with $\omega_n = \beta_n^2\sqrt{EI/(\rho A L^4)}$ in \cref{tab:freq}, all accurate to within $0.03\%$. Free vibration from the first eigenmode ($w_\mathrm{tip}(0) = L/100$, $\Delta t = T_1/200$, ten periods) conserves energy to $\sim 10^{-12}$ (\cref{fig:freq}).

\begin{table}[h]
\centering
\caption{Natural frequency comparison: FEM vs.\ analytical.}
\label{tab:freq}
\begin{tabular}{cccc}
\toprule
Mode & FEM (Hz) & Analytical (Hz) & Relative error \\
\midrule
1 & 20.4498 & 20.4497 & 0.0005\% \\
2 & 128.161 & 128.157 & 0.003\% \\
3 & 358.933 & 358.846 & 0.024\% \\
\bottomrule
\end{tabular}
\end{table}

\begin{figure}[htbp]
\centering
\includegraphics[width=\textwidth]{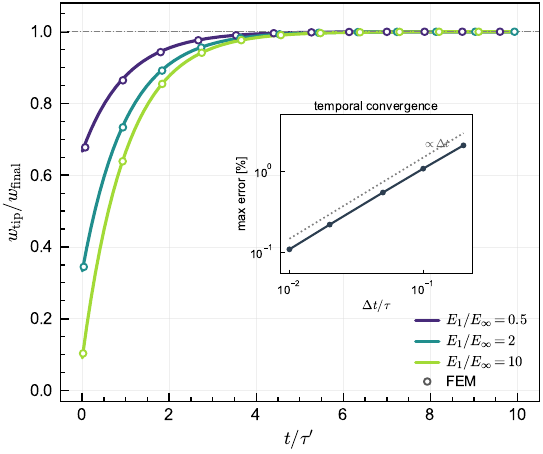}
\caption{SLS creep verification. Solid lines: analytical creep response for three modulus ratios $(E_U - E_R)/E_R$ (color). Open circles: FEM ($\Delta t/\tau_r = 0.1$). The dash-dotted line marks the relaxed ($E_R$) asymptote. Inset: worst-case relative error over the three ratios versus time step, following the first-order ($\propto\Delta t$) reference and remaining below $2\%$ up to $\Delta t/\tau_r = 0.2$.}
\label{fig:creep}
\end{figure}

\paragraph{Standard linear solid} A single-layer SLS beam ($E_R = 1$\,GPa, $\tau_r = 1$\,s) under constant tip load creeps from $w(0^+) = P_0 L^3/(3E_U I)$ to $w(\infty) = P_0 L^3/(3E_R I)$ with time constant $\tau' = \tau_r E_U/E_R$. \Cref{fig:creep} shows agreement with the analytical creep for modulus ratios $(E_U - E_R)/E_R = 0.5,\,2,\,10$. The worst-case error follows the expected first-order slope in the time step and stays below $2\%$ up to $\Delta t/\tau_r = 0.2$.

\begin{figure}[htbp]
\centering
\includegraphics[width=\textwidth]{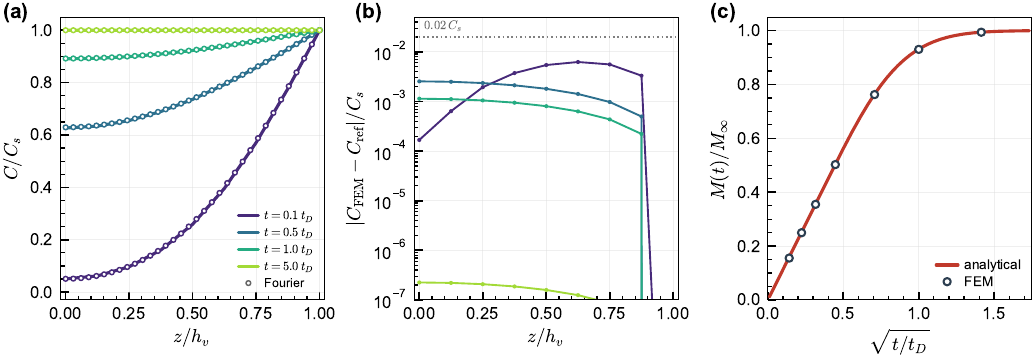}
\caption{Through-thickness diffusion verification at four time instants ($t = 0.1$ to $5\,t_D$). (a)~Concentration profiles: FEM ($N_z = 8$, lines) vs.\ the Fourier series solution (50 terms, circles), indistinguishable at all times. (b)~Absolute error $|C_\mathrm{FEM} - C_\mathrm{ref}|/C_s$, everywhere well below the $0.02\,C_s$ tolerance and localized at the moving front. (c)~Fractional mass uptake $M(t)/M_\infty$ versus $\sqrt{t/t_D}$: the early-time linearity is the Fickian $\sqrt{t}$ signature, and FEM (circles) matches the analytical uptake (line) to within $0.5\%$.}
\label{fig:diffusion}
\end{figure}

\paragraph{Through-thickness diffusion} The one-dimensional diffusion equation (zero initial concentration, zero-flux Neumann at the interface, Dirichlet $C = C_s$ at the surface; $h_1 = 0.1$\,mm, $D = 10^{-12}$\,m$^2$/s) is compared with the Fourier-series solution~\cite{crank1976} at four times in \cref{fig:diffusion}. The absolute error stays below $0.02\,C_s$ and localizes at the moving front. Refining from $N_z = 4$ to $8$ reduces the peak error by roughly a factor of five, consistent with the second-order spatial accuracy of linear elements. Beyond $N_z = 8$ the error saturates near $2.6\times 10^{-3}\,C_s$, limited by the backward-Euler step and the truncation of the reference series. The mass uptake $M(t) = \int_0^{h_1}C\,dz$ matches the analytical result to within $0.5\%$ and shows the Fickian $\sqrt{t}$ signature.

\clearpage
\section{Axial force balance}\label{app:axial}

The beam formulation fixes the reference axis at the $E_R$-weighted neutral axis and omits the uniform axial strain $\varepsilon_0$ (\cref{eq:kinematics}). To bound the error, we compare it with a quasi-static bilayer model on a fine through-thickness grid (200 intervals in the substrate and 400 in the coating) that uses the analytical Fickian concentration and the same exponential update of the standard linear solid. In the full variant, $\varepsilon = \varepsilon_0 + \kappa z$ and both the axial force and the bending moment vanish, with $\varepsilon_0$ and $\kappa$ as unknowns. In the fixed-axis variant, $\varepsilon = \kappa(z - z_R)$ about the $E_R$-weighted axis $z_R$, and only the moment balance is imposed, as in the beam formulation. \Cref{tab:axial} lists the results for the parameters of \cref{sec:crossover}. The steady state is identical, because the coating is fully relaxed and the $E_R$-weighted axis is then the true neutral axis. The overshoot ratios differ by less than $0.3\%$. The largest error occurs for the instantaneous response of an unrelaxed coating, whose stiffer layer shifts the true neutral axis away from $z_R$. It grows with thickness and reaches $3\%$ at $m = 10$. The full-variant steady states also agree with the beam finite-element results to five significant digits.

The thickness study has $E_U h_1/(E_2 h_2) \le 0.1$, whereas the modulus-ratio study of \cref{sec:map} reaches $0.5$. We therefore repeated the comparison over the grid of \cref{fig:map}b, with $E_U/E_R$ from $1.25$ to $10$ and $m$ from $0.2$ to $10$ (\cref{tab:axial_map}). The error grows with the modulus ratio, because the unrelaxed coating then carries a larger share of the axial stiffness. At $E_U/E_R = 10$, the fixed axis overestimates the overshoot ratio by up to $2.7\%$, for thin coatings whose ratio is about $4.5$, and the instantaneous response by $13.5\%$ at $m = 10$. The overshoot threshold is less sensitive and changes by at most $1.1\%$. Near the threshold, uptake is slow compared with relaxation, so the coating is largely relaxed while it loads. The modulus ratio at which the threshold reaches the static optimum is $9.27$ with the fixed axis and $9.52$ with full balance. The beam finite-element model gives $9.25$, close to the fixed-axis bilayer model as expected.

\begin{table}[ht]
\centering
\caption{Fixed-axis formulation compared with full axial and moment balance (parameters of \cref{sec:crossover}). The instantaneous response is the elastic response of the unrelaxed coating ($E_U$) to a uniform eigenstrain.}
\label{tab:axial}
\small
\begin{tabular}{lcccc}
\toprule
 & Steady state & \multicolumn{2}{c}{$\kappa_\mathrm{peak}/\kappa_\mathrm{ss}$} & Instantaneous \\
\cmidrule(lr){3-4}
$m$ & difference & full & fixed axis & difference \\
\midrule
0.2 & $< 0.01\%$ & 1.9207 & 1.9225 & $+0.10\%$ \\
0.5 & $< 0.01\%$ & 1.7144 & 1.7177 & $+0.25\%$ \\
1   & $< 0.01\%$ & 1.3862 & 1.3899 & $+0.47\%$ \\
2   & $< 0.01\%$ & 1.0252 & 1.0262 & $+0.83\%$ \\
4.2 & $< 0.01\%$ & 1.0000 & 1.0000 & $+1.39\%$ \\
10  & $< 0.01\%$ & 1.0000 & 1.0000 & $+3.01\%$ \\
\bottomrule
\end{tabular}
\end{table}

\begin{table}[ht]
\centering
\caption{Fixed-axis formulation compared with full axial and moment balance over the modulus-ratio study of \cref{sec:map} ($m = 0.2$ to $10$, other parameters as in \cref{sec:crossover}). The overshoot-ratio difference is the largest over $m$. The instantaneous difference is given at $m = 10$, where it is largest.}
\label{tab:axial_map}
\small
\begin{tabular}{lccccc}
\toprule
 & $E_U h_1/(E_2 h_2)$ & $\kappa_\mathrm{peak}/\kappa_\mathrm{ss}$ & \multicolumn{2}{c}{$m_\mathrm{os}$} & Instantaneous \\
\cmidrule(lr){4-5}
$E_U/E_R$ & at $m = 10$ & difference & full & fixed axis & difference \\
\midrule
1.25 & 0.06 & $0.05\%$ & 1.323 & 1.326 & $+1.0\%$ \\
1.5  & 0.08 & $0.12\%$ & 1.640 & 1.644 & $+1.8\%$ \\
2    & 0.10 & $0.27\%$ & 2.035 & 2.042 & $+3.0\%$ \\
3    & 0.15 & $0.60\%$ & 2.557 & 2.568 & $+4.8\%$ \\
5    & 0.25 & $1.23\%$ & 3.247 & 3.266 & $+7.6\%$ \\
10   & 0.50 & $2.66\%$ & 4.266 & 4.313 & $+13.5\%$ \\
\bottomrule
\end{tabular}
\end{table}

\section{Static and dynamic response of the coated cantilever}\label{app:dynamic}

\begin{figure}[htbp]
\centering
\includegraphics[width=\textwidth]{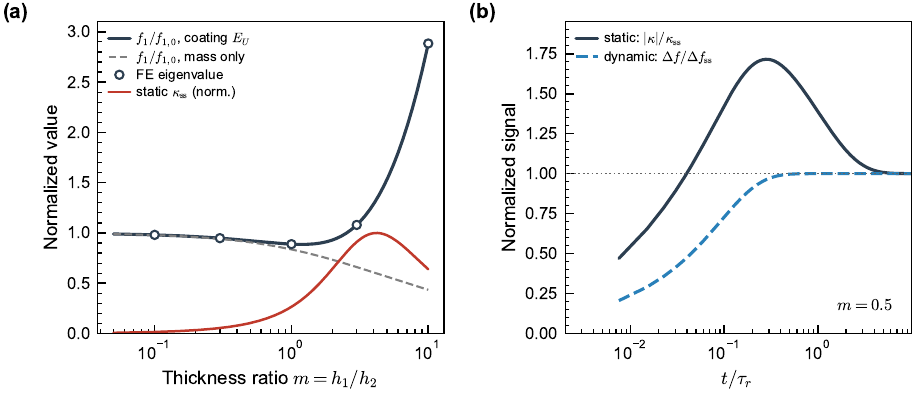}
\caption{Static and dynamic behavior of the coated cantilever. (a)~First bending resonance $f_1$ relative to the bare substrate versus $m$, for a coating with $E_U = 1.7$\,GPa and density $1000$\,kg/m$^3$ on silicon ($h_2 = 3\,\mu$m, $E_2 = 170$\,GPa, density $2330$\,kg/m$^3$). Solid line, composite beam with the unrelaxed coating modulus. Dashed line, coating mass only. Circles, eigenvalues of the beam finite-element model. Red line, steady-state static signal normalized to its maximum. (b)~Transient read-outs for $m = 0.5$ (parameters of \cref{sec:crossover}). Static curvature (solid) and frequency shift from mass loading alone with fixed mechanical properties (dashed), which follows the fractional uptake $M(t)/M_\infty$.}
\label{fig:static_dynamic}
\end{figure}

\Cref{fig:static_dynamic}a shows the first bending resonance of the coated cantilever relative to the bare substrate. At the oscillation frequency the coating responds with its unrelaxed modulus. The composite Euler--Bernoulli value agrees with the eigenvalues of the beam finite-element model to better than $10^{-5}$. The coating mass dominates for thin coatings and lowers the frequency, while the bending stiffness of thick coatings raises it. The static signal, in contrast, peaks at $m_\mathrm{opt}$. \Cref{fig:static_dynamic}b compares the two transient read-outs for one coating. The static curvature overshoots, whereas the frequency shift in the mass-loading-only approximation follows the monotonic uptake.

\end{document}